\documentclass[usenatbib,usegraphicx,useAMS]{mn2e}
\usepackage{amsmath}
\usepackage{amssymb}

\newcommand{\Sg}{Sgr A$^*$}

\def\cf{{cf.}}
\def\eg{{e.g.}}
\def\ie{{i.e.}}

\def\etc{{etc}}

\newcommand{\FigPath}{LowResColorFigs}

\title{Covariant Magnetoionic Theory II: Radiative Transfer}
\author[Avery Broderick \& Roger Blandford]
{Avery Broderick$^1$\thanks{aeb@tapir.caltech.edu}
\& Roger Blandford$^{1,2}$\\
$^1$MC 130-33, Caltech, Pasadena, CA 91125\\
$^2$KIPAC, SLAC, 2575 Sand Hill Road, Menlo Park, CA 94025}

\begin{document}
\maketitle

\begin{abstract}
Accretion onto compact objects plays a central role in high energy
astrophysics. In these environments, both general relativistic and
plasma effects may have significant impacts upon the spectral and
polarimetric properties of the accretion flow.  In paper I we presented
a fully general relativistic magnetoionic theory, capable of tracing
rays in the geometric optics approximation through a magnetised plasma
in the vicinity of a compact object.  In this paper we discuss how to
perform polarised radiative transfer along these rays.  In addition we apply
the formalism to a barotropic thick disk model, appropriate for low
luminosity active galactic nuclei.  We find that it is possible to
generate large fractional polarisations over the innermost portions
of the accretion flow, even when the emission mechanism is unpolarised.
This has implications for accreting systems ranging from pulsars and X-ray
binaries to AGN.
\end{abstract}


\section{Introduction} \label{I}
The spectral and polarimetric properties of astrophysical objects can
provide significant insights into their structure and dynamics.  As a
result, a number of theoretical investigations into the source of
these properties have been undertaken.  Many of these have been
primarily concerned with the spectral properties alone, typically
comparing a physically motivated accretion flow to observations.
However, with the measurement of polarisation in a number of sources,
a significant fraction of the focus has been turned towards
reproducing their polarimetric properties.  In the context of an
accreting compact object, both general relativistic and plasma effects
can play a role in determining these properties.  In \citealp{Brod-Blan:03b}
(hereafter Paper I) we demonstrated how to construct ray trajectories,
in the geometric optics approximation, in a magnetoactive plasma in a
relativistic environment.  In order to apply this to realistic
accretion environments it is necessary to be able to perform radiative
transfer along these rays.

Non-refractive, polarised radiative transfer through magnetised plasmas
is flat space has been extensively studied.  A number of examples
involving weak magnetic fields exist in the literature
\citep[see \eg][]{Sazo-Tsyt:68,Sazo:69,Jone-ODel:77b,Jone-ODel:77a,Ginz:70}.
More recently, investigations into the net effects of tangled
magnetic fields (expected to be typical in magnetised accretion flows)
have begun \citep[see \eg][]{Rusz-Bege:02}.  However, none of these
deal with general relativistic environments.

The importance of refraction in the propagation of radio wavelengths
has long been appreciated in the context of the ionosphere
\citep[see \eg][]{Budd:64,Ginz:70}.  More recently, refraction has
been studied in conjunction with pulsars
\citep[see \eg][]{Welt-Stap-Horn-Edwa:03,Petr:02,Petr:00,
Barn-Aron:86a,Aron-Barn:86b}.  Nonetheless, in all of these cases, the emission
was assumed to originate from a region distinct from where the
refraction occurred.  Refractive lensing of neutron stars was
considered by \citealp{Shav-Heyl-Lith:99}, but ignored general
relativisitic effects.

General relativistic studies into the propagation of polarisation in
vacuum have been done.  These have been primarily interested in the
geometrical effects due to the parallel transport of the linear
polarisation \citep[see \eg][]{Agol:97,Laor-Netz-Pira:90,Conn-Star-Pira:80}.
Alternatively, in \citealp{Brom-Meli-Liu:01},
polarised emission in a general relativistic environment is
considered. However, none of the typical plasma transfer
effects (\eg~Faraday rotation) were included in these calculations.  
In \citealp{Heyl-Shav-Lloy:03}, the vacuum birefringence due to strong
magnetic fields was considered in the context of neutron star atmospheres. 
However, in both, refraction was completely ignored.
There have been some attempts to study the problem of ray propagation
in a covariant form \citep[see \eg][]{Brod-Blan:03b,Geda-Melr:01}, but
in these the radiative transfer was not addressed.

As discussed in Paper I, refraction coupled with the presence of a
horizon can be a source of significant polarisation when the
observation frequency is near the plasma and cyclotron frequencies of
emitting region.  The sense of the resulting net polarisation is determined
by the plasma parameters at the surface at which the polarisation freezes
out (when the modes cease to be adiabatic and must be treated as if
they were in vacuum).  Typically, this will result in a net circular
polarisation.  In a future paper we will discuss astrophysical
environments in which this may be the case, including applications to
\Sg and high mass X-ray binaries.

We present a method for performing polarised radiative transfer
through a strongly refractive magnetised plasma in a general relativistic
environment. Additionally, we apply this to a model of a thick accretion disk.
This is done in six sections with \S\ref{RP} briefly reviewing the
formalism presented in Paper I, \S\ref{PRTiRP} discussing how to
perform the radiative transfer in a magnetised plasma, \S\ref{LHSRiCP}
presenting low harmonic synchrotron as a possible
emission mechanism, \S\ref{R} presenting some results, and \S\ref{C}
containing conclusions.  The details of constructing a magnetised,
thick, barotropic disk are presented in the appendix.

\section{Ray Propagation}
\label{RP}
While astrophysical plasmas will, in general, be hot, the cold case
provides an instructive setting in which to demonstrate the types of
effects that may be present.  As a result, it will be assumed that the
plasma through which the rays propagate will be cold, with a small
component of emitting hot electrons.  As shown in Paper I, the rays
may be explicitly constructed given a dispersion relation,
$D\left( k_\mu, x^\mu \right)$ (a function of the wave four-vector and
position which vanishes along the ray), by integrating the ray equations:
\begin{equation}
\frac{dx^\mu}{d\tau} = \left( \frac{\partial D}{\partial k_\mu} \right)_{x^\mu}
\;\;\mbox{and}\;\;\;
\frac{dk_\mu}{d\tau} = -\left( \frac{\partial D}{\partial x^\mu}
\right)_{k_\mu} \,,
\label{ray_eqs}
\end{equation}
where $\tau$ is an affine parameter along the ray.  Expanding
Maxwell's equations in the geometric optics limit provides the
polarisation eigenmodes and the dispersion relation (given a
conductivity):
\begin{equation}
\left( k^\alpha k_\alpha \delta^\mu_\nu - k^\mu k_\nu
- 4\pi i \omega \sigma^\mu_{~\nu} \right) E^\nu = 0 \,,
\label{disp_eq}
\end{equation}
where $E^\mu$ is the four-vector coincident with the electric field in
the locally flat, comoving rest frame (LFCR frame),
$\omega \equiv -\overline{u}^\mu k_\mu$ ($\overline{u}^\mu$ is the plasma
four-velocity which defines the LFCR frame), and $\sigma^\mu_{~\nu}$
is the covariant extension of the the conductivity tensor.  For the
cold, magnetoactive, electron-ion plasma (in the limit of infinite ion
mass), the dispersion relation is
\begin{multline}
D\left( k_\mu, x^\mu \right)
= \\
k^\mu k_\mu - \delta \omega^2
- \frac{\delta}{2 \left( 1 + \delta \right)}
\Biggl\{ \Biggl[ \left( \frac{e {\cal B}^\mu k_\mu}{m \omega} \right)^2
- \left( 1 + 2 \delta \right) \omega_B^2 \Biggr] \\
\pm
\sqrt{
\left( \frac{e {\cal B}^\mu k_\mu}{m \omega} \right)^4
+ 2 \left( 2 \omega^2 - \omega_B^2 - \omega_P^2 \right)
\left( \frac{e {\cal B}^\mu k_\mu}{m \omega} \right)^2
+ \omega_B^4
}
\Biggr\} \,,
\label{disp_gen}
\end{multline}
where ${\cal B}^\mu$ is the four-vector coincident with external
magnetic field in the LFCR frame, $\omega_P$ is the plasma frequency
in the LFCR frame, $\omega_B$ is the cyclotron frequency associated
with ${\cal B}^\mu$, and
$\delta \equiv \omega_P^2/(\omega_B^2 - \omega_P^2)$,
This is a covariant form of the Appleton-Hartree dispersion
relation \citep[see \eg][]{Boyd-Sand:69}.

In general, the electromagnetic polarisation eigenmodes will not follow the
same trajectories, and in particular will not follow null geodesics.
As a result, the different polarisation eigenmodes will sample
different portions of the accretion flow.  As shown in Paper I, it is
possible for one mode to be captured by the central black
hole while the other escapes, leading to a net polarisation.

\section{Polarised Radiative Transfer in Refractive Plasmas}
\label{PRTiRP}
Both emission and absorption are local processes.  However, because
the transfer of radiation necessarily involves a comparison between
the state of the radiation at different points in space, global
propagation effects need to be accounted for.  These take two general
forms: correcting for the gravitational redshift; and keeping track of
the local coordinate system, \ie~ensuring that polarised emission is
being added appropriately in the presence of a rotation of the
coordinate system propagated along the ray.  In addition, for a
magnetoactive plasma, it is necessary to determine how to perform the
radiative transfer in the presence of refraction.

\subsection{Length Scales and Regimes}
\label{PRTiRP:LSaR}
The problem of performing radiative transfer in a magnetoactive plasma
has been treated in detail in the context of radio-wave propagation in
the ionosphere \citep[for a detailed discussion see
\eg][]{Ginz:70,Budd:64}. In these cases it was found that there were
two distinct limiting regimes.  These can be distinguished by
comparing two fundamental scales of the affine parameter $\tau$: that
over which the polarisation eigenmodes change appreciably, $\tau_S$,
and the Faraday rotation length, $\tau_F$.  Before $\tau_S$ can be
defined it is necessary to define a pair of basis four-vectors that
define the axes of the ellipse:
\begin{align}
\hat{e}^\mu_{\parallel} &= 
\frac{
\left( k^\alpha k_\alpha + \omega^2 \right) {\cal B}^\mu
-
{\cal B}^\nu k_\nu \left( k^\mu - \omega \overline{u}^\mu \right)
}
{
\sqrt{ k^\beta k_\beta + \omega^2 }
\sqrt{
\left( k^\sigma k_\sigma + \omega^2 \right)
{\cal B}^\gamma {\cal B}_\gamma
-
\left({\cal B}^\gamma k_\gamma\right)^2
}
}
\\
\hat{e}^\mu_{\perp} &=
\frac{
\varepsilon^{\mu\nu\alpha\beta} \overline{u}_\nu k_\alpha {\cal B}_\beta
}
{
\sqrt{
\left( k^\sigma k_\sigma + \omega^2 \right) {\cal B}^\gamma {\cal B}_\gamma
-
\left( {\cal B}^\gamma k_\gamma \right)^2
}
}
\,,
\label{tetrad_defs}
\end{align}
where $\varepsilon^{\mu\nu\alpha\beta}$ is the Levi-Civita pseudo-tensor.
In terms of these, the ellipticity angle $\chi$ can be defined by
\begin{equation}
\tan \chi \equiv
i
\frac{
e_\parallel^\mu E_{O\,\mu}
}{
e_\perp^\nu  E_{O\,\nu}
}
= 
i
\frac{
e_\perp^\mu E_{X\,\mu}
}{
e_\parallel^\nu E_{X\,\nu}
}
 \,.
\end{equation}
In general, an additional angle, $\phi$, is necessary to define the
polarisation, namely the angle which defines the orientation of the
ellipse.  The basis four-vectors have been chosen such that $\phi$ is
identically zero.  However, this choice introduces a new geometric
term into the equations which accounts for the necessary rotation of
the basis four-vectors, contributing a non-zero $d \phi/d \tau$ (see
\S \ref{PRTiRP:SCR} for more details).
Then, in general,
\begin{equation}
\tau_S \equiv \left( \left| \frac{d \phi}{d \tau} \right|^2 + 
\left| \frac{d \chi}{d \tau} \right|^2 \right)^{-1/2} \,,
\end{equation}
For the ordered fields employed here (see the
appendices),
\begin{equation}
\tau_S
\simeq \left| \frac{\omega_B}{\omega^3}
\frac{\partial \omega_P^2}{\partial x^\mu}
\frac{d x^\mu}{d \tau} \right|^{-1} \,,
\end{equation}
where this approximation form is true for small cyclotron
and plasma frequencies and all but the most oblique angles of
incidence.  The Faraday rotation length is defined to be the distance
over which the phase difference between the two polarisation
eigenmodes reaches $2 \pi$, \ie
\begin{equation}
\tau_F \equiv \left| \Delta k_\mu \frac{d x^\mu}{d \tau} \right|^{-1} \,,
\end{equation}
where $\Delta k^\mu$ is the difference between the wave vectors of the
two modes.  Strictly speaking in addition to $\tau_F$, $\tau_S$ should
be compared to a term describing the rate of change of the Faraday rotation
length, however in the situations under consideration here this term is
completely dominated by $\tau_F$.

Together, these length scales define three regimes: the
{\em adiabatic} regime ($\tau_F \ll \tau_S$), the {\em intermediate}
regime ($2 \tau_F \sim \tau_S$), and the {\em strongly coupled} regime
($\tau_F \gg \tau_S$).  In all regimes the polarisation of the plasma
eigenmodes is uniquely set by the dispersion equation, equation
(\ref{disp_eq}).

In general, as $\theta\rightarrow\pi/2$, $\Delta k \simeq
(\omega_P^2 \omega_B/ \omega^2 c) \cos \theta + (\omega_P^2
\omega_B^2/\omega^3 c)$, where $\theta$ is the angle between the
wave-vector and the magnetic field.  Hence to remain in the adiabatic regime
$\tau_S \gg (\omega/\omega_B)^2 \tau_F(\theta=0)$, which is typically
not true in astrophysical sources.  As a result, as the magnetic field
becomes perpendicular to the wave-vector, the modes generally become
strongly coupled.  This is the reason why, when dealing with a large
number of field reversals (\eg~in a molecular cloud), the amount of
Faraday rotation and conversion is
$\propto {\bf B} \cdot d{\bf x}$ and not $ | {\bf B} | \cdot d{\bf x}$
(which would follow in the adiabatic regime)
despite the fact that $\tau_s \gg \tau_F(\theta=0)$
may be true throughout the entire region.

\subsection{Adiabatic Regime}
\label{PRTiRP:AR}
In the adiabatic regime the two polarisation modes
propagate independently \citep[see \eg][]{Ginz:70}.  As a result,
to a good approximation, the polarisation is simply given by the
sum of the two polarisations.
The intensities, $I_O$ and $I_X$, of the ordinary and the
extraordinary modes, respectively, are not conserved along the ray due
to the gravitational redshift.  Consequently, the photon occupation
numbers of the two modes, $N_O$ and $N_X$, which are Lorentz scalars,
and hence are conserved along the rays, are used.  Therefore, the
equation of radiative transfer is given by
\begin{equation}
\frac{d N_{O,X}}{d \tau} = \frac{d l}{d \tau} \left(
\overline{j}_{O,X} - \alpha_{O,X} N_{O,X} \right) \,,
\end{equation}
where
\begin{equation}
\frac{d l}{d \tau} = \sqrt{
g_{\mu\nu} \frac{d x^\mu}{d \tau} \frac{d x^\nu}{d \tau}
-
\left( u_\mu \frac{d x^\mu}{d \tau} \right)^2}
\end{equation}
is the conversion from the line element in the LFCR frame to the affine
parameterisation, and $\overline{j}_{O,X}$ is the emissivity in the
LFCR frame scaled appropriately for the occupation number (as opposed
to the intensity).  In practice, the occupation numbers will be large.
However, up to fundamental physical constants, it is permissible to
use a scaled version of the occupation numbers such that
$N_{O,X} = \omega^{-3} I_{O,X}$ in vacuum.

It is also this regime in which Faraday rotation and conversion occur.
However, because these propagation effects result directly from
interference between the two modes, and
hence require the emission to be coherent among the two modes, when
they diverge sufficiently the modes must be added incoherently and
thus Faraday rotation and conversion effectively cease.  The modes
will have divereged sufficiently when
\begin{equation}
|\Delta x_\perp | \gtrsim \frac{\lambda^2}{\Delta \lambda} \,,
\label{FR_cohe_limit}
\end{equation}
where $\Delta \lambda$ is the emission band-width.  For continum
emission, this reduces to $|\Delta x_\perp| \gtrsim \lambda$. Therefore
in a highly refractive medium an additional constraint is placed upon
Faraday rotation.  The depth at which equation (\ref{FR_cohe_limit})
is first satisfied can be estimated by considering an oblique ray
entering a  plane-parallel density and magnetic field distribution (at
angle $\zeta$ to the gradient).
In this case, to linear order in $\omega_P$ and $\omega_B$,
\begin{equation}
\frac{d^2 \Delta x_\perp}{d z^2}
\simeq
- \sin\zeta \frac{\partial D}{\partial z}
\simeq
\frac{\omega_B \omega_P^2}{\omega^3 z} \\
\end{equation}
As a result,
\begin{equation}
| \Delta x_\perp | \simeq \frac{\omega_B \omega_P^2 z}{2 \omega^3},
\quad {\rm hence} \quad
z_{\rm max} \simeq \sqrt{ \lambda \frac{2 \omega^3}{\omega_B \omega_P^2}} \,.
\end{equation}
The resulting number of Faraday rotations, $n_F$, is then given by,
\begin{equation}
n_F \equiv \int_0^{z_{\rm max}} \frac{\Delta k}{2 \pi} dz
\simeq \frac{1}{2 \pi \sin \zeta} \,,
\end{equation}
which is typically small for all but the smallest $\zeta$.  Because, as
discussed in section \ref{R}, linear polarisation is strongly
suppressed by refraction, such a small Faraday rotation in negligible.
As a result, for the situations of interest here, in this regime the
modes can be added together incoherently to yield the net polarisation.

\subsection{Strongly Coupled Regime}
\label{PRTiRP:SCR}
In the limit of vanishing plasma density it is clear that the polarisation
propagation must approach that in vacuum regardless of the magnetic field
geometry.  In this limit the two modes must be strongly coupled such that
their sum evolves as in vacuum.  In particular, it is necessary to
keep track of their relative phases.  This can be most easily accomplished
by using the Stokes parameters to describe the radiation.  In this case also
it is possible to account for the gravitational redshift by using the
photon occupation number instead of intensities, $N$, $N_Q$, $N_U$,
$N_V$.  However, it is also necessary to define the $N_Q$, $N_U$, and
$N_V$ in a manner that is consistent along the entire ray.  In order
to do this we may align the axes of $N_Q$ along the magnetic field, \ie
\begin{align}
N_Q &= N(\hat{e}^\mu_{\parallel}) - N(\hat{e}^\mu_{\perp}) \nonumber \\
N_U &= N\left( \frac{1}{\sqrt2}\hat{e}^\mu_{\parallel}
-\frac{1}{\sqrt2} \hat{e}^\mu_{\perp} \right)
- N\left(\frac{1}{\sqrt2} \hat{e}^\mu_{\parallel}
+ \frac{1}{\sqrt2}\hat{e}^\mu_{\perp} \right) \\
N_V &= N\left(\frac{1}{\sqrt2} \hat{e}^\mu_{\parallel}
+ \frac{i}{\sqrt2}\hat{e}^\mu_{\perp}\right)
- N\left(\frac{1}{\sqrt2} \hat{e}^\mu_{\parallel}
- \frac{i}{\sqrt2}\hat{e}^\mu_{\perp}\right) \,, \nonumber
\end{align}
where $N(e^\mu)$ is the occupation number of photons in the polarisation
defined by $e^\mu$.  Thus the problem of relating $N_Q$, $N_U$, and
$N_V$ along the ray is reduced to propagating
$\hat{e}^\mu_{\parallel}$ and $\hat{e}^\mu_{\perp}$.  A change in
$\tau$ by $d \tau$ is associated with a rotation of the basis by an
angle 
\begin{equation}
d \phi =
\hat{e}_{\perp \mu} \frac{d x^\nu}{d \tau} \nabla_\nu \hat{e}_{\parallel}^\mu d\tau
\,,
\end{equation}
where the use of the covariant derivative, $\nabla_\nu$, accounts for
the general relativistic rotations of $\hat{e}_{\parallel}^\mu$ and
$\hat{e}_{\perp}^\mu$.
As a result, the transfer effect due to general relativity {\em and}
the rotation of the magnetic field about the propagation path is
\begin{align}
\frac{d N_Q}{d\tau} &= -2 \frac{d \phi}{d \tau} N_U \nonumber \\
\frac{d N_U}{d\tau} &= 2 \frac{d \phi}{d \tau} N_Q \,,
\end{align}
where the factor of 2 arises from the quadratic nature of N.

After a specific emission model is chosen the emissivities and the
absorption coefficients are scaled as in \S\ref{PRTiRP:AR}.  An example will
be discussed in more detail in \S\ref{LHSRiCP}.

\subsection{Intermediate Regime}
\label{PRTiRP:IR}
At some point it is necessary to transition from one limiting regime to
the other.  In this intermediate regime the polarisation freezes out.  A
great deal of effort has been expended to understand the details of
how this occurs \citep[see \eg][]{Budd:52}.  However, to a good
approximation it is enough to set the polarisation at the point
when $\tau_F = 2\tau_S$ to the incoherent sum of the polarisation eigenmodes
\citep[see the discussion in][]{Ginz:70}:
\begin{align}
N &= N_O + N_X  \nonumber \\
N_Q &= -\cos 2\chi (N_O - N_X) \nonumber \\
N_U &= 0 \\
N_V &= \sin 2\chi (N_O - N_X) \nonumber
\end{align}
It is straightforward to show that in terms of the generalised Stokes
parameters $N_O$ and $N_X$ are given by (this is true even when they
are offset by a phase)
\begin{align}
N_O &= \frac12 \left( N - \cos 2\chi N_Q + \sin 2\chi N_V \right)
\nonumber \\
N_X &= \frac12 \left( N + \cos 2\chi N_Q - \sin 2\chi N_V \right) \,.
\label{Stokes_proj}
\end{align}
Note that, in general, polarisation information will be lost in this
conversion.  This is a reflection of the fact that the space
spanned by the incoherent sum of the two modes forms a subset of the
space of unpolarised Stokes parameters.  This is clear from their
respective dimensionalities; the former is three
dimensional (there are only three degrees of freedom for the
decomposition into the two polarisation modes, namely their amplitudes
and relative phase), while the later is four dimensional ($I$, $Q$,
$U$, and $V$, subject only to the condition that $I^2 \ge Q^2+U^2+V^2$).

\section{Low Harmonic Synchrotron Radiation into Cold Plasma Modes}
\label{LHSRiCP}
As discussed in the previous section, emission and absorption are
inherently local processes.  As a result it will be sufficient in this
context to treat them in the LFCR frame, and hence in flat space.  In
this frame it is enough to solve the problem in three dimensions and
then insert quantities in a covariant form.

Because refractive effects become large only when $\omega \sim
\omega_B, \omega_P$, for there to be significant spectral and
polarimetric effects it is necessary to have an emission mechanism
which operates in this frequency regime as well.  A plausible
candidate is low harmonic synchrotron emission.  It is assumed that a
hot power-law distribution of electrons is responsible for the
emission while the cold plasma is responsible for the remaining plasma
effects.  In Paper I we did present the theory for the warm plasma as
well, however, as in the conventional magnetoionic theory, it is much
more cumbersome to utilise.

\subsection{Razin Suppression}
\label{LHSRiCP:RS}
A well known plasma effect upon synchrotron emission is the Razin suppression
\citep[see \eg][]{Rybi-Ligh:79, Beke:66}.
This arises due to the increase in the wave phase velocity above the speed
of light, preventing electrons from maintaining phase with the emitted
electromagnetic wave, resulting in an exponential suppression of the emission
below the Razin frequency,
\begin{equation}
\omega_R = \frac{\omega_P^2}{\omega_B} \,.
\end{equation}
However, as discussed in the Appendix, for the disk model
we have employed here, typically $\omega_B>\omega_P$ and hence the
Razin effects do not arise.

\subsection{Projection onto Non-Orthogonal Modes}
\label{LHSRiCP:PoNOM}
A significant problem with emission mechanisms in the $\omega \sim
\omega_B, \omega_P$ frequency regime is that the modes are no longer
orthogonal.  It is true that for a lossless medium (such as the
cold plasma), equation (\ref{disp_eq}), which defines the
polarisation, is self-adjoint.  However, because of the $k^\mu$ differ
for the two modes, it is a slightly different equation for each mode,
and hence the polarisations are eigenvectors of slightly different
hermitian differential operators.  In the high frequency limit this difference
becomes insignificant.

The energy in the electromagnetic portion of the wave (neglecting the
plasma portion) is given by
\begin{equation}
{\cal E} = \frac{{\bf E}^* \cdot \bmath{\epsilon} \cdot {\bf E}}{4 \pi}
= \frac{1}{4 \pi} {\bf E}^* \cdot
\left( \bmath{1} + \frac{4 \pi i}{\omega}\bmath{\sigma} \right) \cdot {\bf E}
\end{equation}
For each mode (${\bf E}_O$ and ${\bf E}_X$), the dispersion equation gives
\begin{align}
\left( \omega^2 + 4 \pi i \omega \bmath{\sigma} \right) \cdot {\bf E}_{O,X}
&= \left( k_{O,X}^2 - {\bf k}_{O,X} \otimes {\bf k}_{O,X} \right)
\cdot {\bf E}_{O,X} \nonumber \\
&= k_{O,X}^2 \left( {\bf 1} - {\bf \hat{k}} \otimes {\bf \hat{k}}
\right) \cdot {\bf E}_{O,X} \,.
\end{align}
Therefore, with ${\bf E} = \sum_i {\bf E}_i$, 
\begin{equation}
{\cal E} = \frac{1}{4 \pi \omega^2} \sum_{i,j} k_j^2 {\bf E}_i^* \cdot 
\left( {\bf 1} - {\bf \hat{k}} \otimes {\bf \hat{k}} \right) \cdot
{\bf E}_j \,.
\end{equation}
However, for a lossless medium it is also true that
\begin{equation}
{\cal E} = {\cal E}^\dagger
= \frac{1}{4 \pi \omega^2} \sum_{i,j} k_i^2 {\bf E}_i^* \cdot
\left( {\bf 1} - {\bf \hat{k}} \otimes {\bf \hat{k}} \right) \cdot
{\bf E}_j \,,
\end{equation}
and therefore,
\begin{equation}
 \sum_{i,j} \left( k_i^2 - k_j^2 \right) {\bf E}_i^* \cdot
\left( {\bf 1} - {\bf \hat{k}} \otimes {\bf \hat{k}} \right)
\cdot {\bf E}_j = 0 \,.
\end{equation}
For a non-degenerate dispersion relation, \eg~that of a magnetoactive
plasma, this implies that the the components of the polarisation
transverse to the direction of propagation are orthogonal for the two
modes, \ie
\begin{equation}
{\bf \hat{F}}_i^* \cdot {\bf \hat{F}}_j = k_i^2 \delta_{ij}
\end{equation}
where
\begin{equation}
{\bf \hat{F}}_{O,X} =  k_{O,X}
\frac{ \left( {\bf 1}- {\bf \hat{k}} \otimes {\bf \hat{k}} \right)
\cdot {\bf \hat{E}}_{O,X}}
{{\bf \hat{E}}_{O,X}^* \cdot
\left( {\bf 1}- {\bf \hat{k}} \otimes {\bf \hat{k}} \right)
\cdot {\bf \hat{E}}_{O,X}}  \,.
\end{equation}
As a result it is possible to define ${\cal E}_{O,X}$ such that
\begin{equation}
{\cal E}_{O,X} = \frac{{\bf F}_{O,X}^* \cdot {\bf F}_{O,X}}{4 \pi}
\quad {\rm and} \quad
{\cal E} = \sum_i {\cal E}_i \,,
\label{energy_decomp}
\end{equation}
\ie~that the electromagnetic energy can be uniquely decomposed into the
electromagnetic energy in the two modes.

Expressions for the ${\bf F}_{O,X}$ can be obtained by solving for the
eigenvectors of the dispersion equation.  For the cold magnetoactive
plasma this gives
\begin{multline}
{\bf \hat{F}}_{O,X} =
\frac{k_{O,X}}{\sqrt2} \left[ 
\sqrt{ 1 \pm \left( 1+ \varepsilon \right)^{-1/2}} \, {\bf
  \hat{e}}_\parallel \right.\\
\left.
\pm \; i \sqrt{ 1 \mp \left( 1+ \varepsilon \right)^{-1/2}} \, {\bf
  \hat{e}}_\perp \right]
\,,
\end{multline}
where, (not to be confused with the Levi-Civita pseudo-tensor)
\begin{equation}
\varepsilon = \left( \frac{\sin^2 \theta}{2 \cos \theta}
\frac{\omega \omega_B}{\omega_P^2 - \omega^2} 
\right)^{-2} \,,
\end{equation}
$\theta$ is the angle between the magnetic field and the wave
vector, and ${\bf \hat{e}}_{\parallel,\perp}$ are the flat space
analogues of the basis vectors in equation (\ref{tetrad_defs}).
$\theta$ may be defined covariantly by 
\begin{equation}
\cos^2 \theta = \frac{\left({\cal B}^\mu k_\mu\right)^2}
{{\cal B}^\nu {\cal B}_\nu \left( k^\sigma k_\sigma + \omega^2
  \right)}
\,.
\end{equation}
This corresponds to the polarisation found in the literature
\citep[\cf][]{Budd:64}.

\subsection{Emissivities}
\label{LHSRiCP:E}
Because the energy can be uniquely decomposed into the energy in each
polarisation eigenmode, it is possible to calculate the emissivities
and absorption coefficients by the standard far-field method.  For
synchrotron radiation this was originally done by \citealp{West:59}.
The calculation is somewhat involved but straightforward and has
been done in detail in the subsequent literature
\citep[see \eg][]{Rybi-Ligh:79}.  Consequently, only the result for the
power emitted (per unit frequency and solid angle) for a given
polarisation is quoted below:
\begin{multline}
\langle P^{O,X}_{\omega\,\Omega} \rangle
=
\frac{e^3 {\cal B} \sin \theta}{8 \sqrt3 \pi^2 m k_{O,X}^2}
n_r^2
\int d^3\!p f({\bf p}) \Bigg[ \\
\left(
\left| {\bf \hat{F}}_{O,X} \cdot {\bf \hat{e}}_\parallel \right|^2 
+
\left| {\bf \hat{F}}_{O,X} \cdot {\bf \hat{e}}_\perp \right|^2
\right) F(x) \\
+\left(
\left| {\bf \hat{F}}_{O,X} \cdot {\bf \hat{e}}_\parallel \right|^2 
-
\left| {\bf \hat{F}}_{O,X} \cdot {\bf \hat{e}}_\perp \right|^2
\right) G(x)
\Bigg] \,,
\label{pow_in_vec}
\end{multline}
where
\begin{equation}
x=\frac{2 m c \omega}{3 \gamma^2 e {\cal B} \sin \theta}\,,
\end{equation}
$f({\bf p})$ is the distribution function of emitting electrons,
$n_r$ is the ray-refractive index
\citep[for a suitable definition see][]{Beke:66},
and $F$ and $G$ have their usual definitions,
\begin{equation}
F(x) = x \int_x^\infty K_{\frac53}(y) dy
\quad {\rm and} \quad
G(x) = x K_{\frac23}(x) \,,
\end{equation}
where the $K_{5/3}$ and $K_{2/3}$ are the modified Bessel functions of
$5/3$ and $2/3$ order, respectively.  The addition factor of $n_r^2$ arises
from the difference in the photon phase space, $d^3\!k$ and the analogous
integral over frequency, $4\pi d\omega$.

For the adiabatic regime, the emissivities, $\overline{j}_{O,X\,\omega}$, can
now be defined:
\begin{equation}
\overline{j}_{O,X} = \frac{1}{4 \pi n_r^2 \omega^3}
\langle P^{O,X}_{\omega\,\Omega} \rangle \,.
\end{equation}
For a power-law distribution of emitting electrons,
$f({\bf p}) d^3\!p = {\cal C} \gamma^{-s} d\gamma$, this gives
\begin{multline}
\overline{j}_{O,X} =
\frac{\sqrt3 e^2{\cal C}}{24 \pi^2 \omega^2 c (1+s)}
\left( 3 \frac{\omega_B}{\omega} \sin\theta \right)^{\frac{s+1}{2}}
\Gamma\left(\frac{s}{4}+\frac{19}{12}\right)
\\ \times
\Gamma\left(\frac{s}{4}-\frac{1}{12}\right)
\left[
1
\pm\;
\frac{3s+3}{3s+7}
\left(1+\varepsilon\right)^{-\frac12}
\right] \,.
\label{emis_adi}
\end{multline}
The Stokes emissivities and absorption coefficients for an emitting hot power
law (ignoring effects of order $\gamma^{-1}$ as these explicitly
involve the propagation through the hot electrons) are given by
\begin{align}
\overline{j}_{N} &=
\overline{j}_{O} + \overline{j}_{X} \\
\overline{j}_{Q} &= 
\frac{\sqrt3 e^2{\cal C}}{48 \pi^2 \omega^2 c}
\left( 3 \frac{\omega_B}{\omega} \sin\theta \right)^{\frac{s+1}{2}}
\nonumber \\
&\quad \times
\Gamma\left(\frac{s}{4}+\frac{7}{12}\right)
\Gamma\left(\frac{s}{4}-\frac{1}{12}\right) \\
\overline{j}_{U} &= \overline{j}_{V} = 0 \,.
\end{align}
Note that for low $\gamma$ synchrotron can efectively produce circular
polarisation, namely $\overline{j}_{V} \sim 3/\gamma$.  The production
of circular polarisation in this way in environments with large
Faraday depths will be considered in future publications.

\subsection{Absorption Coefficients}
\label{LHSRiCP:AC}
For the adiabatic regime, detailed balance for each mode requires
that the absorption coefficients are then given by
\begin{multline}
\alpha_{O,X} =
\frac{\sqrt3 \pi e^2{\cal C}}{6 \omega m c}
\left( 3 \frac{\omega_B}{\omega} \sin\theta \right)^{\frac{s+2}{2}}
\Gamma\left(\frac{s}{4}+\frac{11}{6}\right)
\\ \times
\Gamma\left(\frac{s}{4}+\frac{1}{6}\right)
\left[
1
\pm\;
\frac{3s+6}{3s+10}
\left(1+\varepsilon\right)^{-\frac12}
\right] \,.
\end{multline}

In the strongly coupled regime, the Stokes absorption coefficient
matrix \citep[see \eg][and references therein]{Jone-ODel:77a},
\begin{equation}
\left(
\begin{array}{cccc}
\alpha_N & \alpha_Q & 0 & \alpha_V \\
\alpha_Q & \alpha_N & 0 & 0 \\
0 & 0 & \alpha_N & 0 \\
\alpha_V & 0 & 0 & \alpha_N
\end{array}
\right) \,.
\label{pol_abs}
\end{equation}
where the Faraday rotation and conversion due to the hot electrons have
been ignored as a result of the fact that they will be negligible in
comparison to the Faraday rotation and conversion due to the cold electrons.
The individual $\alpha$'s can be obtained in terms of the
$\alpha_{O,X}$ using the fact that the energy in the electromagnetic
oscillations can be uniquely decomposed into contributions from each
mode (equation (\ref{energy_decomp})).  Then,
\begin{align}
\frac{d N}{d \lambda}
&= \frac{d N_O}{d \lambda} + \frac{d N_X}{d \lambda} \nonumber \\
&= j_O + j_X - \alpha_O N_O - \alpha_X N_X \nonumber \\
&= \left( j_O+j_X \right)
- \frac12 (\alpha_O+\alpha_X) N \\
&\quad + \frac12 \cos 2\chi \left(\alpha_O-\alpha_X\right) Q
- \frac12 \sin 2\chi \left(\alpha_O-\alpha_X\right) V \nonumber \,.
\end{align}
Therefore, the absorption coefficients may be identified as,
\begin{align}
\alpha_N &= \frac12 \left( \alpha_O+\alpha_X \right) \\
\alpha_Q &= -\frac12 \cos 2\chi \left(\alpha_O-\alpha_X\right) \\
\alpha_V &= \frac12 \sin 2\chi \left(\alpha_O-\alpha_X\right) \,.
\end{align}

\subsection{Unpolarised Low Harmonic Synchrotron Radiation}
\label{LHSRiCP:ULHSR}
To highlight the role of refraction in the generation of
polarisation, an unpolarised emission mechanism is also used.  To compare
with the results of the polarised emission model discussed in
the previous section, the artificial scenario in which the synchrotron
emission is split evenly into the two modes was chosen.  In this case,
\begin{equation}
\overline{j}^{\rm UP}_{O,X} =
\frac12 \overline{j}_N \,,
\end{equation}
and
\begin{equation}
\overline{j}^{\rm UP}_{N} = \overline{j}_N \,,
\end{equation}
with the other Stokes emissivities vanishing.  Similarly, the absorption
coefficients are given by,
\begin{equation}
\alpha^{\rm UP}_{O,X} = \alpha^{\rm UP}_N = \alpha_N\,,
\end{equation}
with the other absorption coefficients vanishing as well.

\subsection{Constraints Upon the Emitting Electron Fraction}
For refractive plasma effects to impact the spectral and
polarimetric properties of an accretion flow, it is necessary that it
be optically thin.  This places a severe constraint upon the fraction of
hot electrons, $f \equiv {\cal C}/ [n_e (s-1)]$.  In terms of the plasma
frequency and $f$ the absorptivity is approximately
\begin{equation}
\alpha_N \sim
\frac{\sqrt3}{24c} 
f \frac{\omega_P^2}{\omega}
\left( 3 \frac{\omega_B}{\omega} \sin \theta \right)^{(s+2)/2} \,. 
\end{equation}
With $s\sim 2$, and $\omega \sim \omega_P,\,\omega_B$, the typical optical
depth (not to be confused with the affine parameter) is
\begin{equation}
\tau \sim 10^{-1} f \frac{R}{\lambda}
\quad\quad
{\rm hence}
\quad\quad
f \sim 10 \frac{\lambda}{R} \,,
\end{equation}
where $R$ is the typical disk scale length (here on the order of $10 M$).

\section{Results}
\label{R}

\subsection{Disk Model}
\label{R:DM}
Before any quantitative results are presented it is necessary to select a
specific plasma and magnetic field distribution.  Here this takes the form
of an azimuthally symmetric, thick, barotropic disk around a maximally
rotating Kerr black hole ($a\simeq0.98$).  The magnetic field
is chosen to lie upon surfaces of constant angular velocity, thus insuring
that it does not shear.  In order to maintain such a field it must also be
strong enough to suppress the magneto-rotational instability.  Further
details may be found in the appendix.

\subsection{Ray Trajectories}
\label{R:RT}

\begin{figure*}
\begin{center}
\begin{tabular}{cc}
\includegraphics[width=\columnwidth]{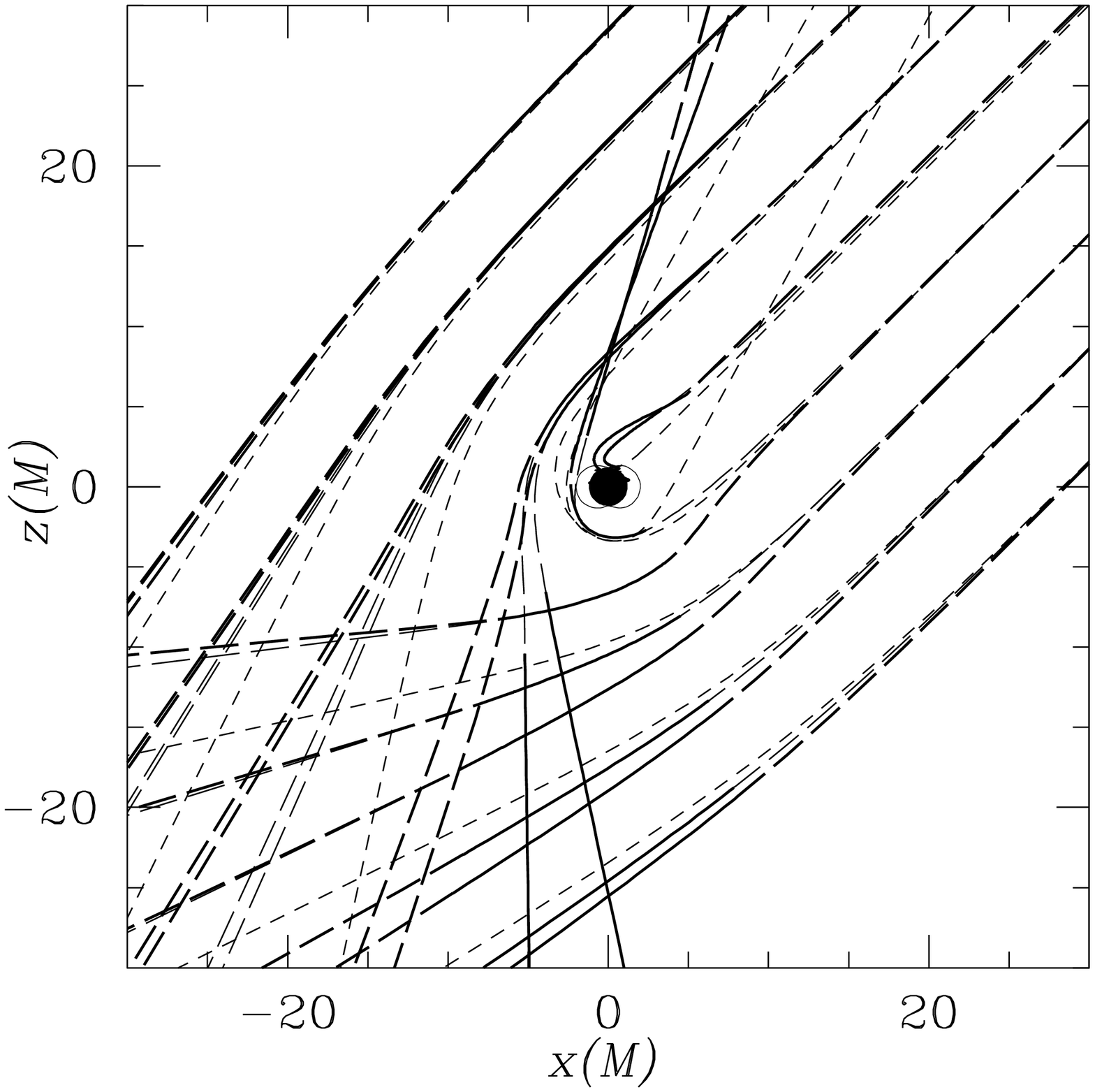}
&
\includegraphics[width=\columnwidth]{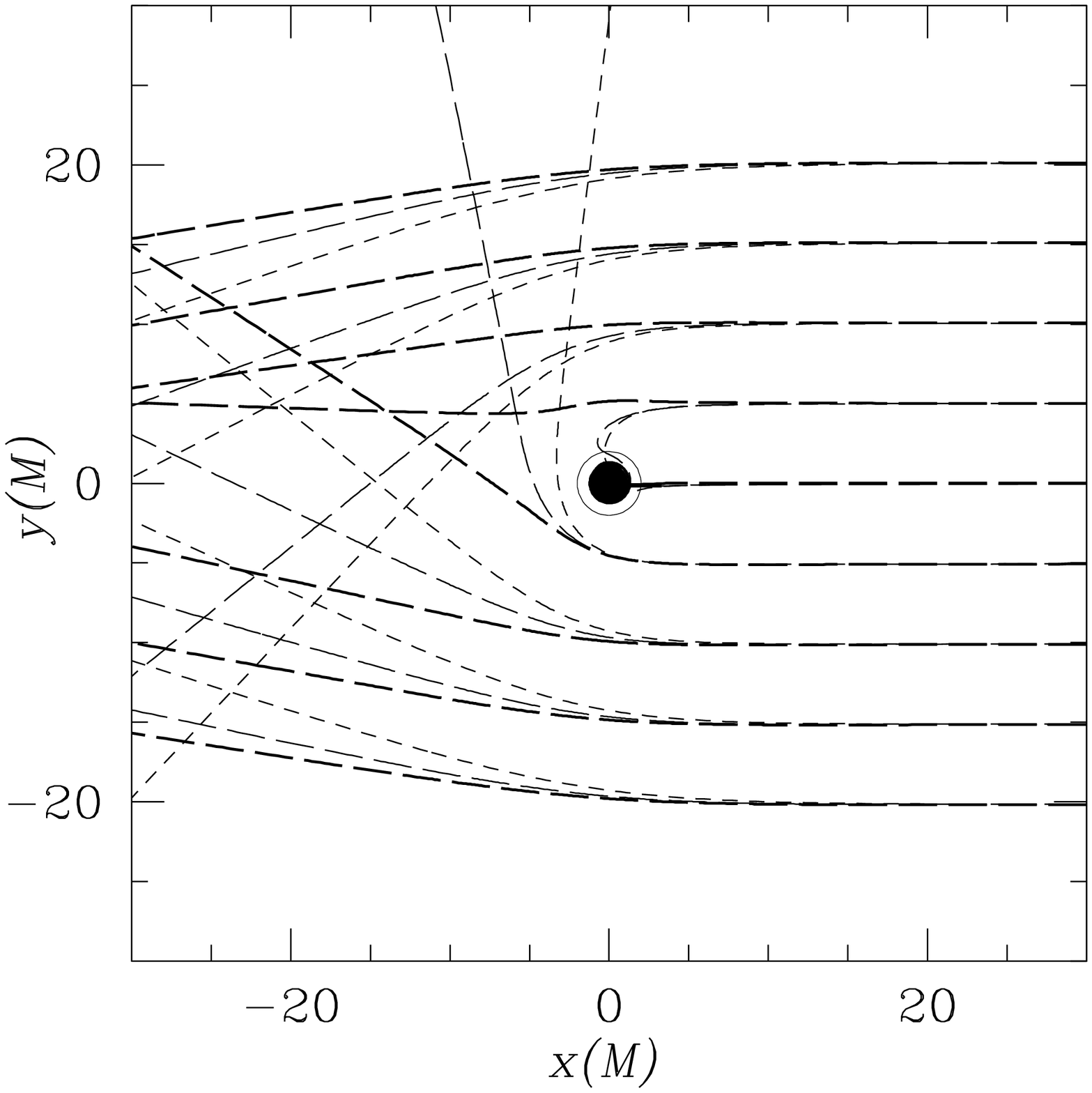}
\\
{\bf (a)}
&
{\bf (b)}
\end{tabular}
\end{center}
\caption{
Shown in panels (a) and (b) are vertical and horizontal cross sections
of rays propagating backwards from an observer located $45^\circ$
above the equatorial plane.  The strongly coupled (adiabatic) regime
is denoted by the solid (long-dashed) lines for the  ordinary (thin)
and extraordinary (thick) polarisation eigenmodes.  For reference, the
null geodesics are drawn in the short dash.  In addition, the black
hole horizon and the boundary of the ergosphere are also shown.
}
\label{vert_horiz_cs}
\end{figure*}

Figure \ref{vert_horiz_cs} shows vertical and horizontal slices
of rays propagated back through the disk discussed in the previous section
from an observer elevated to $45^\circ$ above the equatorial plane at a
frequency $\omega_\infty = 3 \omega_{P\,{\rm max}} /4$.  Note that since
the maximum occurs at $r_{\rm eq}=2M$, the relativistically blue-shifted
$\omega$ is approximately $1.8 \omega_{P\,{\rm max}}$ placing it comfortably
above the plasma resonance at all points (assuming Doppler effects do not
dominate at this point.)

The refractive effects of the plasma are immediately
evident with the extraordinary mode being refracted more so 
\citep[see discussion in][]{Brod-Blan:03b}.  Gravitational lensing is also
shown to be important over a significant range of impact parameters.
There will be an azimuthal asymmetry in the ray paths due to both the 
black hole spin and the Dopper shift resulting from the rotation of the
disk.  This can be clearly observed in panel (b) Figure \ref{vert_horiz_cs}.

In panel (a) of Figure \ref{vert_horiz_cs} the transition between the two
radiative transfer regimes is also clearly demonstrated.  Each time a ray
passes from the strongly coupled to the adiabatic regime it must be
reprojected into the two polarisation eigenmodes.  If the
plasma properties (\eg~density, magnetic field strength or
direction, \etc) are not identical to when the polarisation had previously
frozen out (if at all), this decomposition will necessarily be different.
As a result, when propagating the rays backwards, whenever one passes from the
adiabatic to the strongly coupled regime, it is necessary to follow {\em both}
polarisation eigenmodes in order to ensure the correctness of the radiative
transfer.  The leads to a doubling of the rays at such points.  When
integrating the radiative transfer equations forward along the ray, the
net intensity is then projected out using equation (\ref{Stokes_proj}).
This ray doubling is clearly present in panel (a) of Figure
\ref{vert_horiz_cs}, where the rays pass into the strongly coupled
regime and back again as they traverse the evacuated funnel above and
below the black hole.

Note that the trajectories of the rays depend upon $\omega_P/\omega_\infty$ and
$\omega_B/\omega_\infty$ only (given a specified disk and magnetic
field structure, of course), where $\omega_\infty$ is $\omega$ as
measured at infinity.  Therefore, the paths shown in Figure
\ref{vert_horiz_cs} are valid for any density
normalisation of the disk described in the appendix as long as
$\omega$ is adjusted accordingly.

\subsection{Polarisation Maps}
\label{R:PM}
In order to demonstrate the formalism described in this paper,
polarisation maps were computed for the disk model described
in section \ref{R:DM} and the appendix \ref{aTDM} orbiting a maximally
rotating black hole as seen by an observer at infinity elevated to
$45^\circ$ above the equatorial plane.  Each map shows Stokes $I$, $Q$,
$U$, and $V$.  

As with the rays trajectories, the particular form of the polarisation
maps only depend upon a few unitless parameters.  These necessarily
include $\omega_{P\,{\rm max}}/\omega$ and
$\omega_{B\,{\rm max}}/\omega$ as these define the ray
trajectories.  In addition, the relative brightness depends upon the
optical depth which is proportional to
$(\omega_{P\,{\rm max}}/\omega)^2 (\omega_{B\,{\rm max}}/\omega) M f \omega/c$.
As a result if the following dimensionless quantities remain unchanged,
the polarisation maps shown in the following sections will apply (up to a
constant scale factor)
\begin{align}
\frac{\omega_{P\,{\rm max}}}{\omega_\infty} &= \frac43 \nonumber\\
\frac{\omega_{B\,{\rm max}}}{\omega_\infty} &= \frac43 \nonumber\\
f \frac{M}{\lambda} &= 2.30 \times 10^3 \,.
\label{holonomic_consts}
\end{align}

Despite the fact that the form of the polarisation maps will remain unchanged
if the quantities in equation (\ref{holonomic_consts}) remain constant, the
normalisation will change by a multiplicative constant in the same way
as the source function, namely proportional to $\omega^2_\infty$.
However, an additional multiplicative factor arises from the solid angle
subtended by the source on the sky.  As a result, Stokes $I$, $Q$, $U$,
and $V$ are all shown in units of
\begin{equation}
\left( \frac{M}{D} \right)^2 \! m_e \, \omega_{P\,{\rm max}}^2 \,,
\label{iquv_normalisation}
\end{equation}
where $D$ is the distance to the source.  This amounts to plotting
\begin{equation}
\frac{k T_B}{m_e c^2} \left( \frac{\omega_\infty}{\omega_{P\,{\rm max}}} \right)^2 \,,
\end{equation}
where $T_B$ is the brightness temperature of the source.

\subsubsection{Unpolarised Emission}
\label{R:PM:UE}
\begin{figure*}
\begin{center}
\begin{tabular}{cc}
\includegraphics[width=\columnwidth]{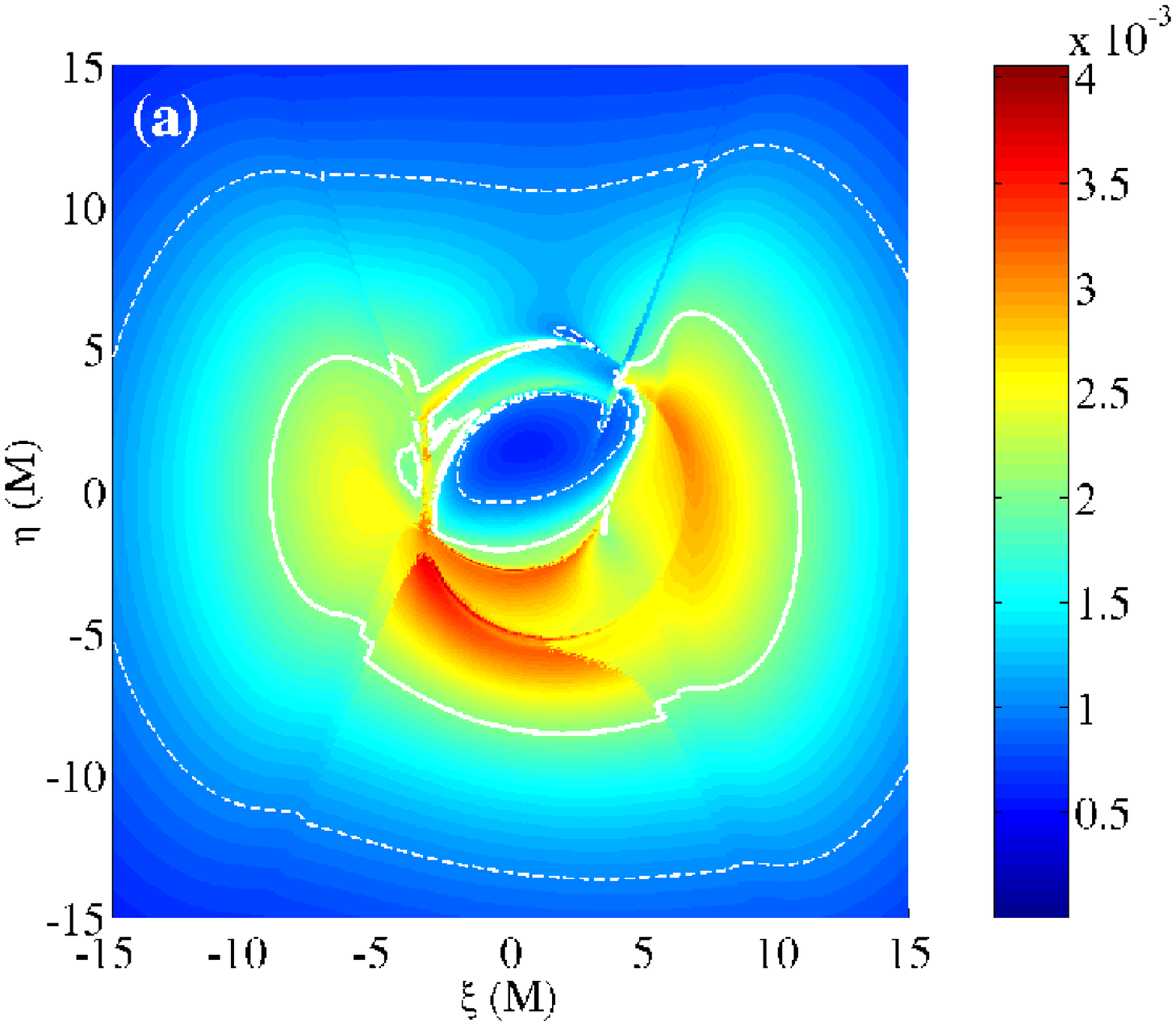}
&
\includegraphics[width=\columnwidth]{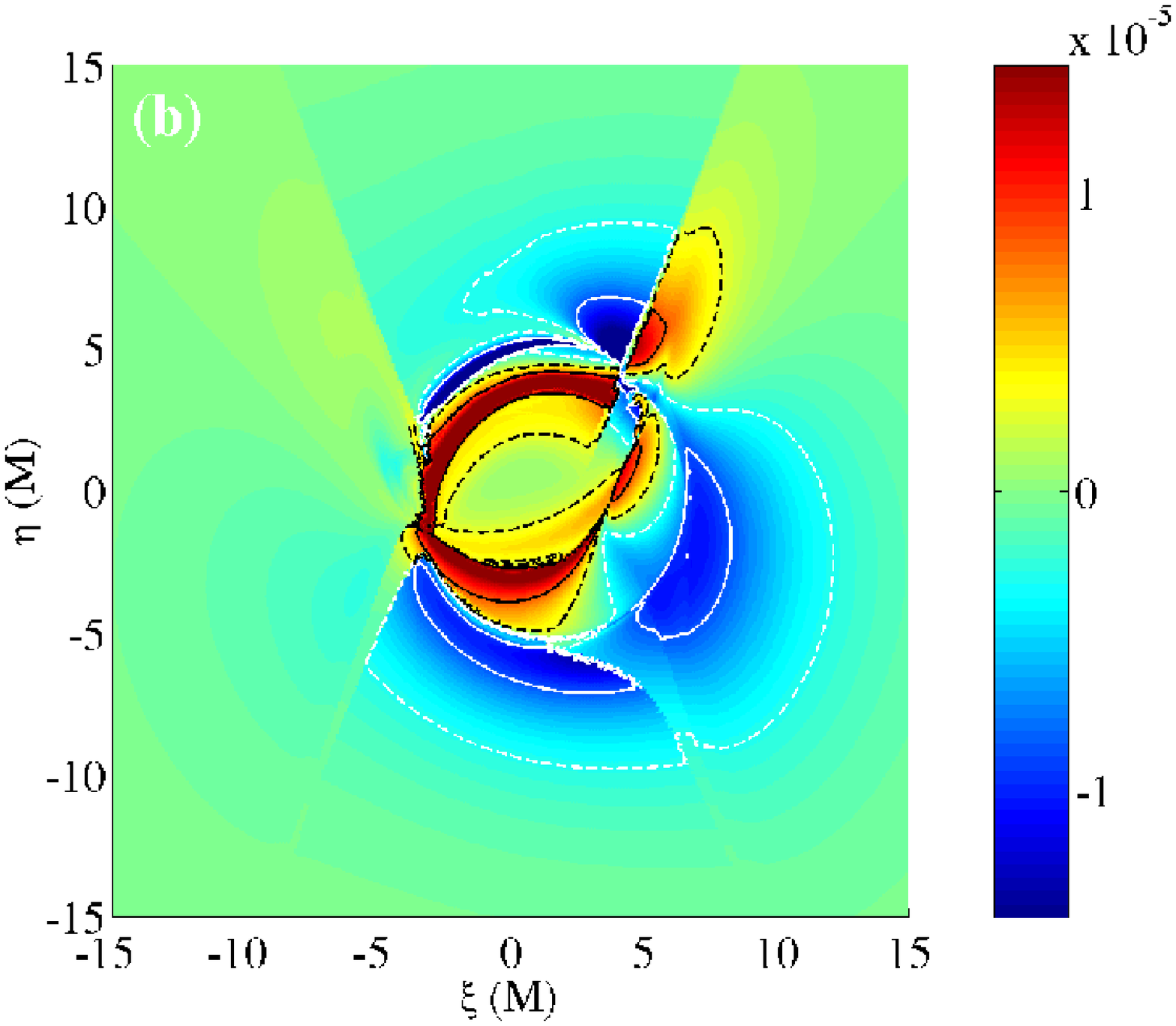} \\
\includegraphics[width=\columnwidth]{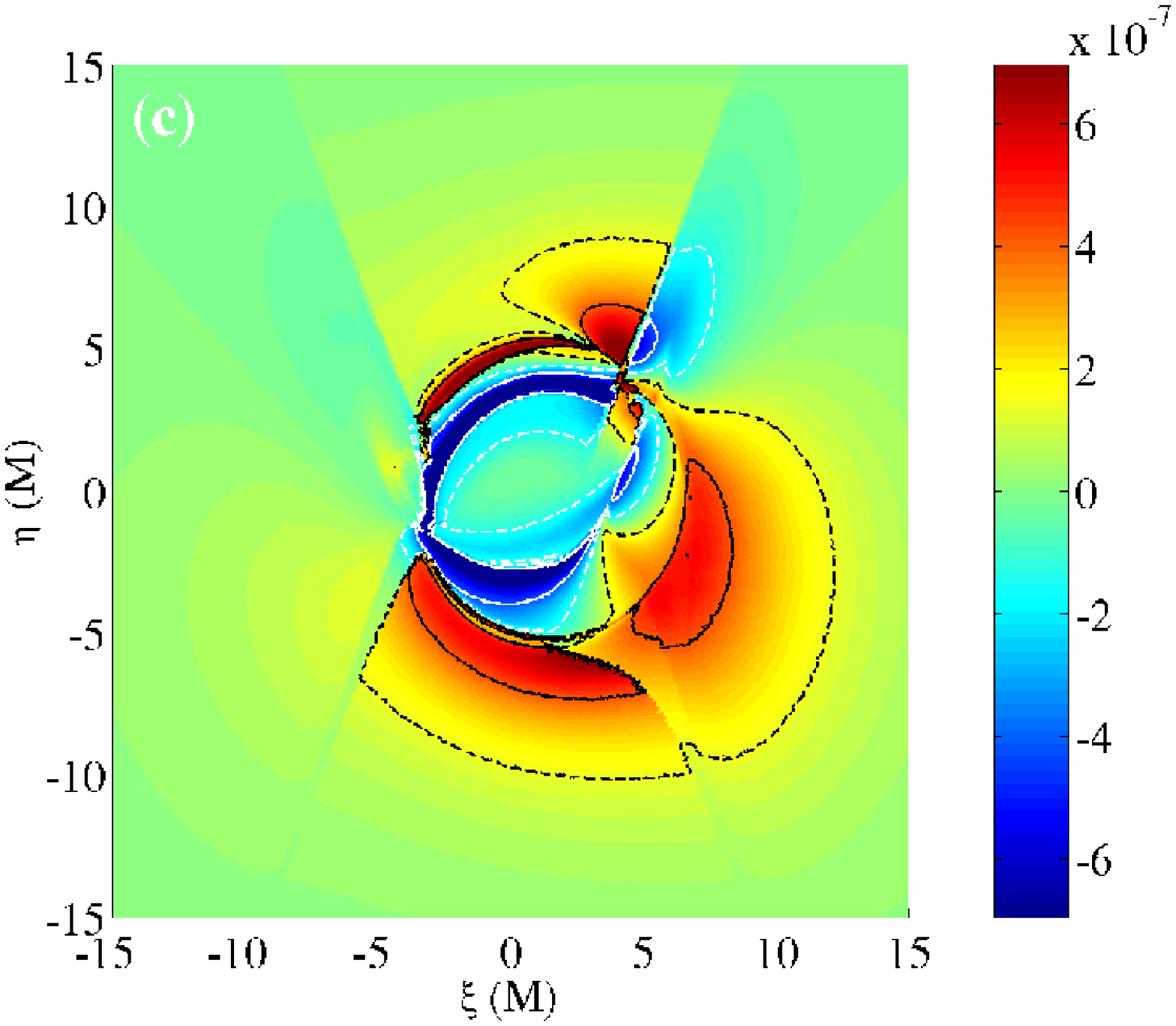}
&
\includegraphics[width=\columnwidth]{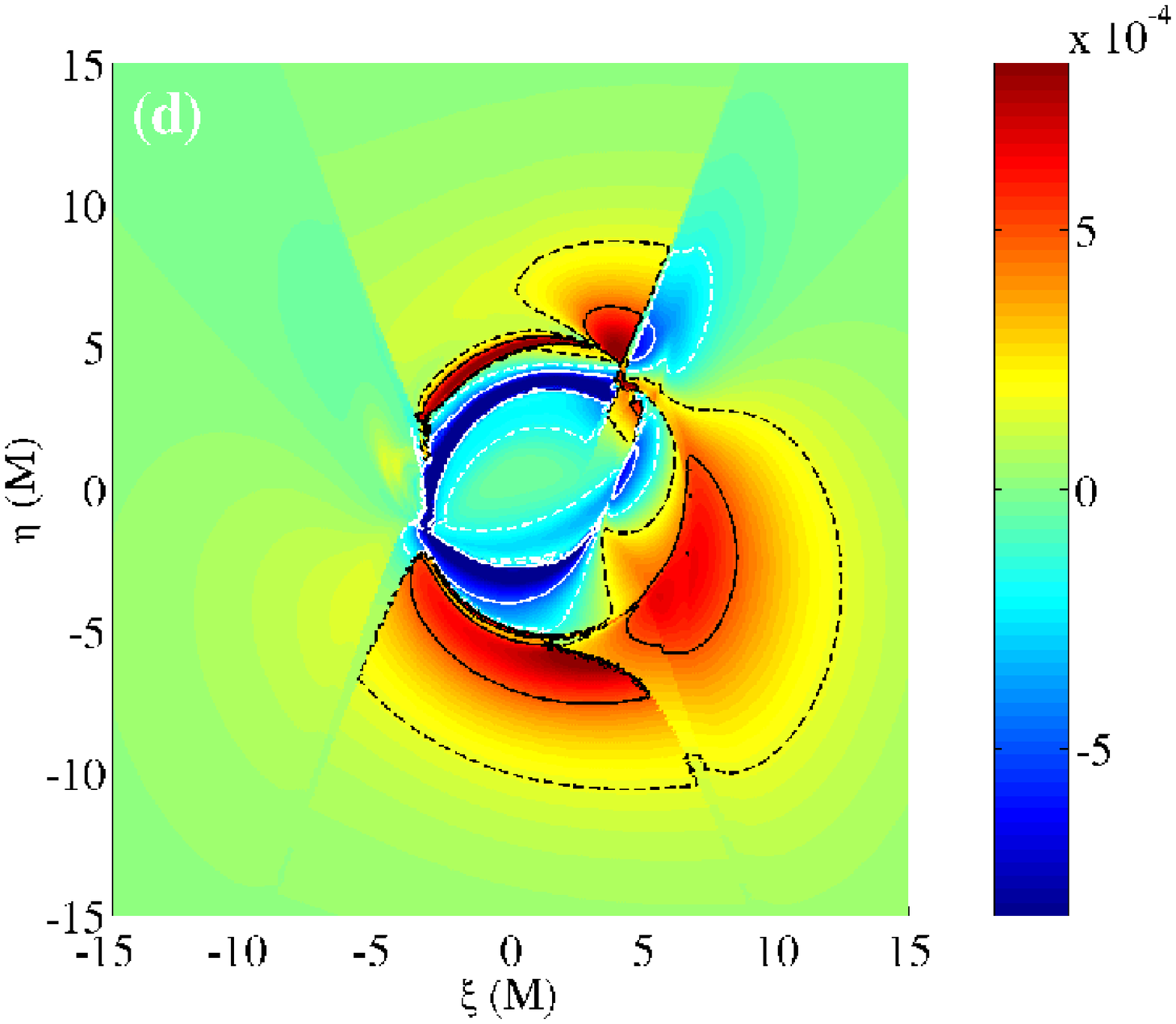}
\end{tabular}
\end{center}
\caption{
Stokes $I$, $Q$, $U$, and $V$ per unit $M^2$ are shown in panels
(a), (b), (c), and (d), respectively, for the unpolarised emission
mechanism described in
section \ref{LHSRiCP:ULHSR} and the disk model described in section
\ref{R:DM} and appendix \ref{aTDM} orbiting a maximally rotating black
hole from a vantage point $45^\circ$ above the equatorial plane at the
frequency $\omega_\infty=3\omega_{P,{\rm max}}/4$.
The contour levels are at 0.2 (dashed) and 0.6 (solid) of the maximum
values shown on the associated colorbars.  The integrated fluxes over the
region shown are $I=1.3$, $Q=-9.4\times10^{-4}$, $U=4.9\times10^{-5}$,
and $V=6.2\times10^{-2}$.  All fluxes are in units of
$(M/D)^{-2} \omega_{P\,{\rm max}}^2$ as discussed above equation
(\ref{iquv_normalisation}).
}
\label{unpol_iquv}
\end{figure*}

For the purpose of highlighting the role of refractive plasma effects
in the production of significant quantities of circular polarisation,
Figure \ref{unpol_iquv} shows Stokes $I$, $Q$, $U$, and $V$ at
$\omega_\infty = 3\omega_{P\,{\rm max}} /4$, calculated using the unpolarised
emission model described in section \ref{LHSRiCP:ULHSR}.
Immediately noticeable are the regions of considerable polarisation
surrounding the black hole.  In addition, the outlines of the evacuated
funnel above and below the hole are clearly visible.

Differences in refraction of the two polarisation eigenmodes leads two
two generic effects: ({\em i}) the presence of two maxima in the intensity
map, each associated with the intensity maxima in a given polarisation
eigenmode; and ({\em ii}) a net excess of one polarisation, and in particular,
circular polarisation.  The polarisation changes rapidly at the 
edges of the evacuated funnels because the refraction and mode decomposition
changes rapidly for modes that just enter the funnel and those that pass
wide of it.  Note that {\em all} of the polarisation is due {\em entirely}
to refractive plasma effects in this case.  The integrated values for
the Stokes parameters are $I=1.3$, $Q=-9.4\times10^{-4}$,
$U=4.9\times10^{-5}$, and $V=6.2\times10^{-2}$, demonstrating that
there does indeed exist a significant net circular polarisation.

\begin{figure*}
\begin{center}
\begin{tabular}{cc}
\includegraphics[width=\columnwidth]{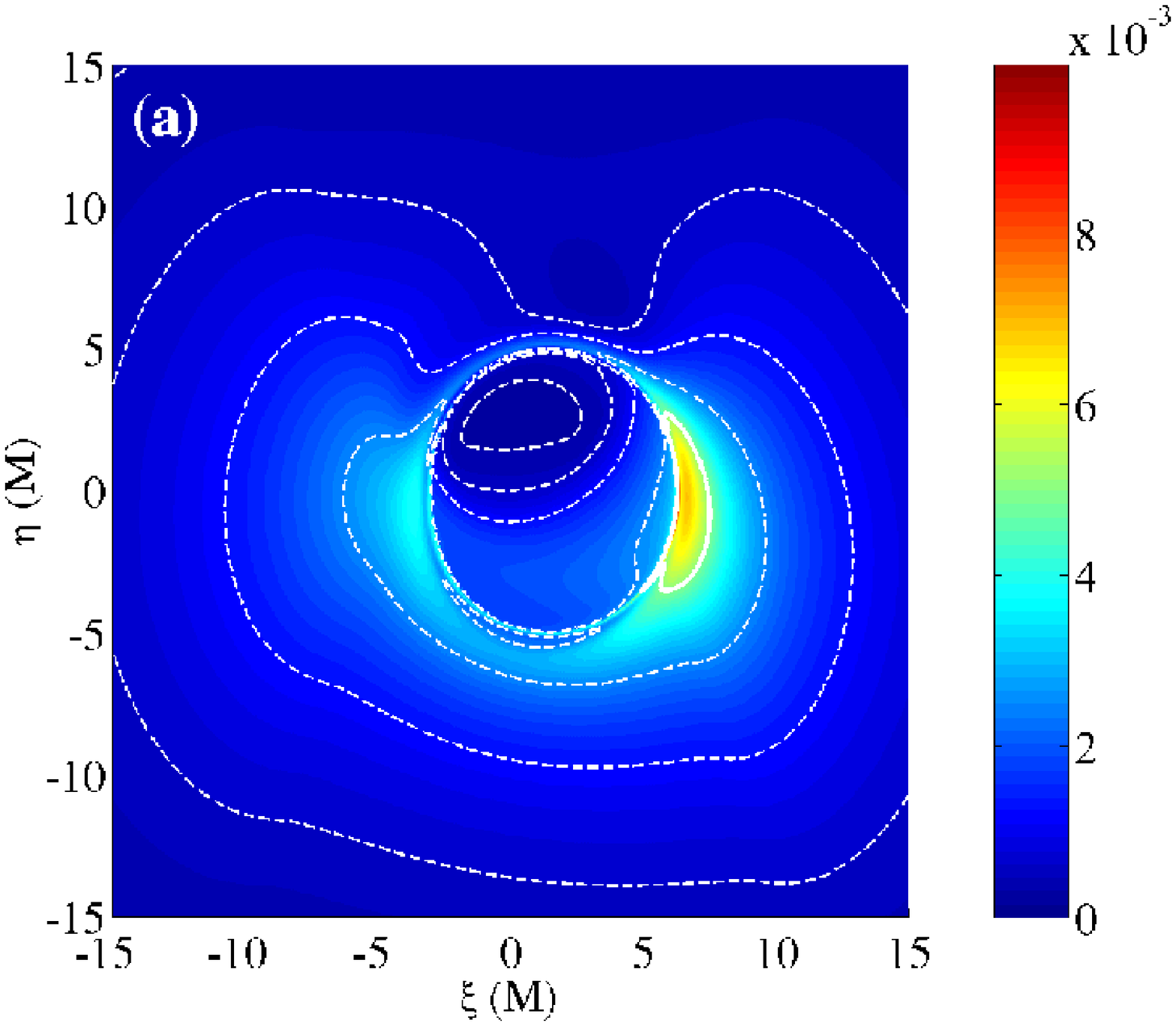}
&
\includegraphics[width=\columnwidth]{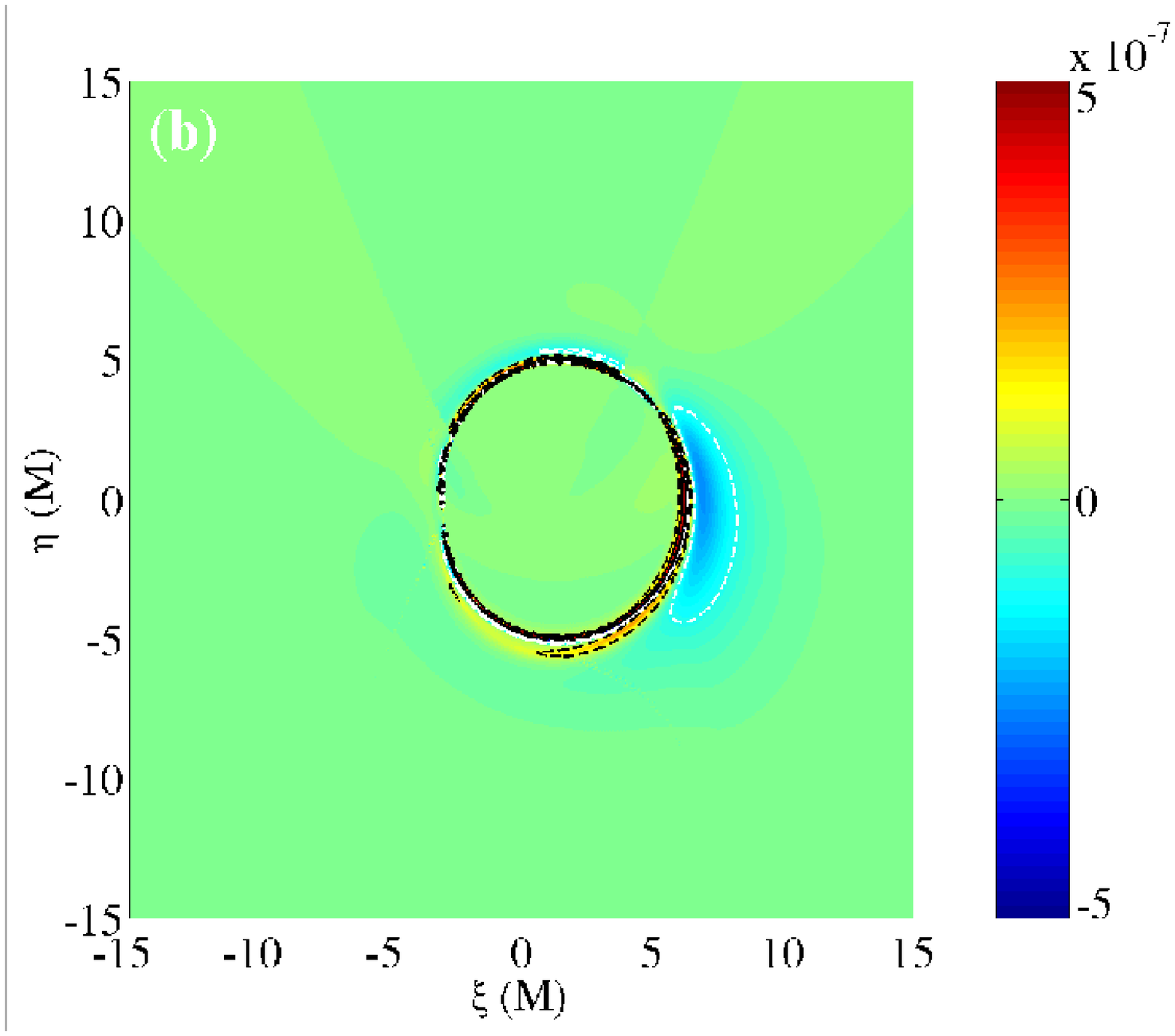} \\
\includegraphics[width=\columnwidth]{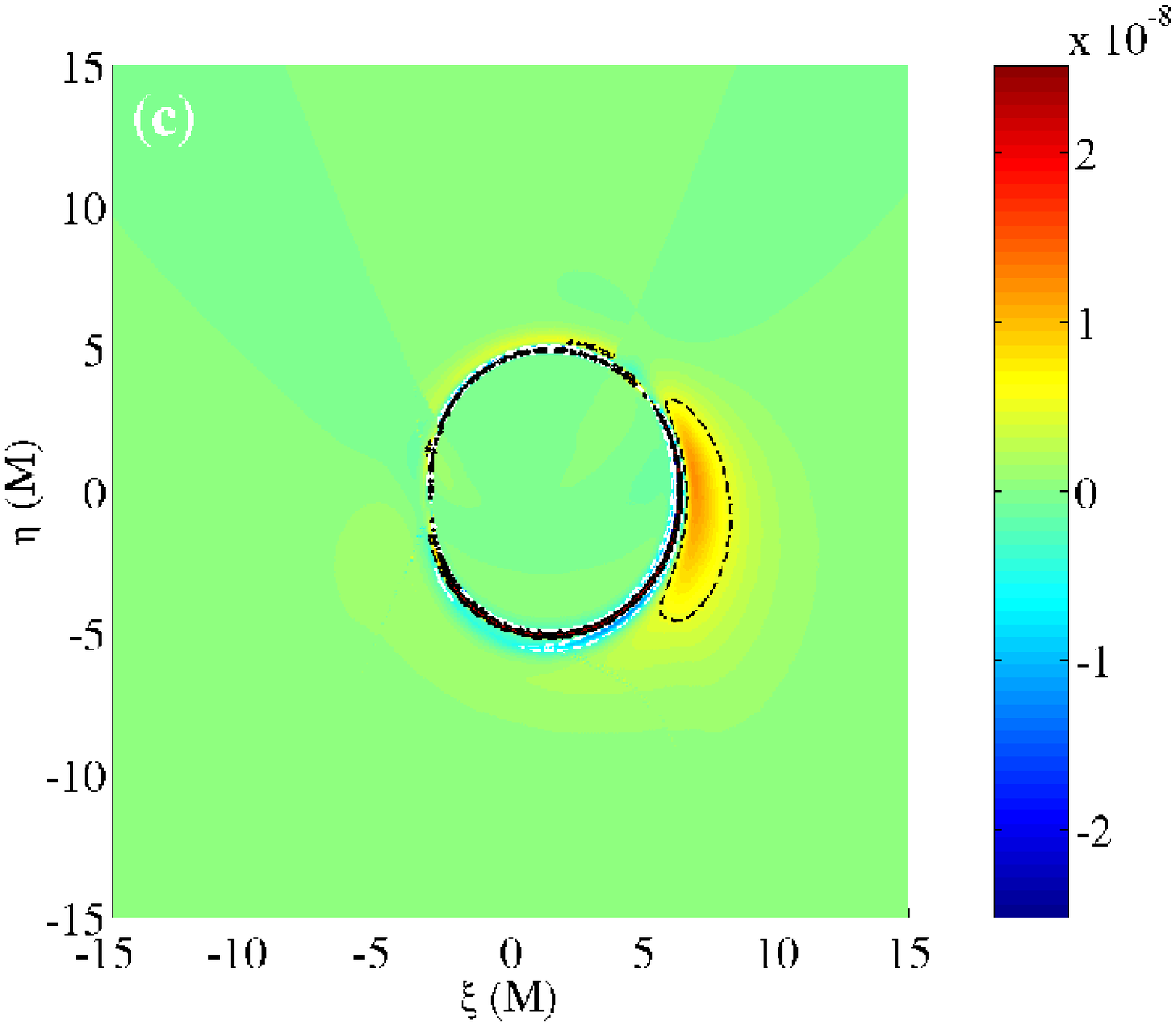}
&
\includegraphics[width=\columnwidth]{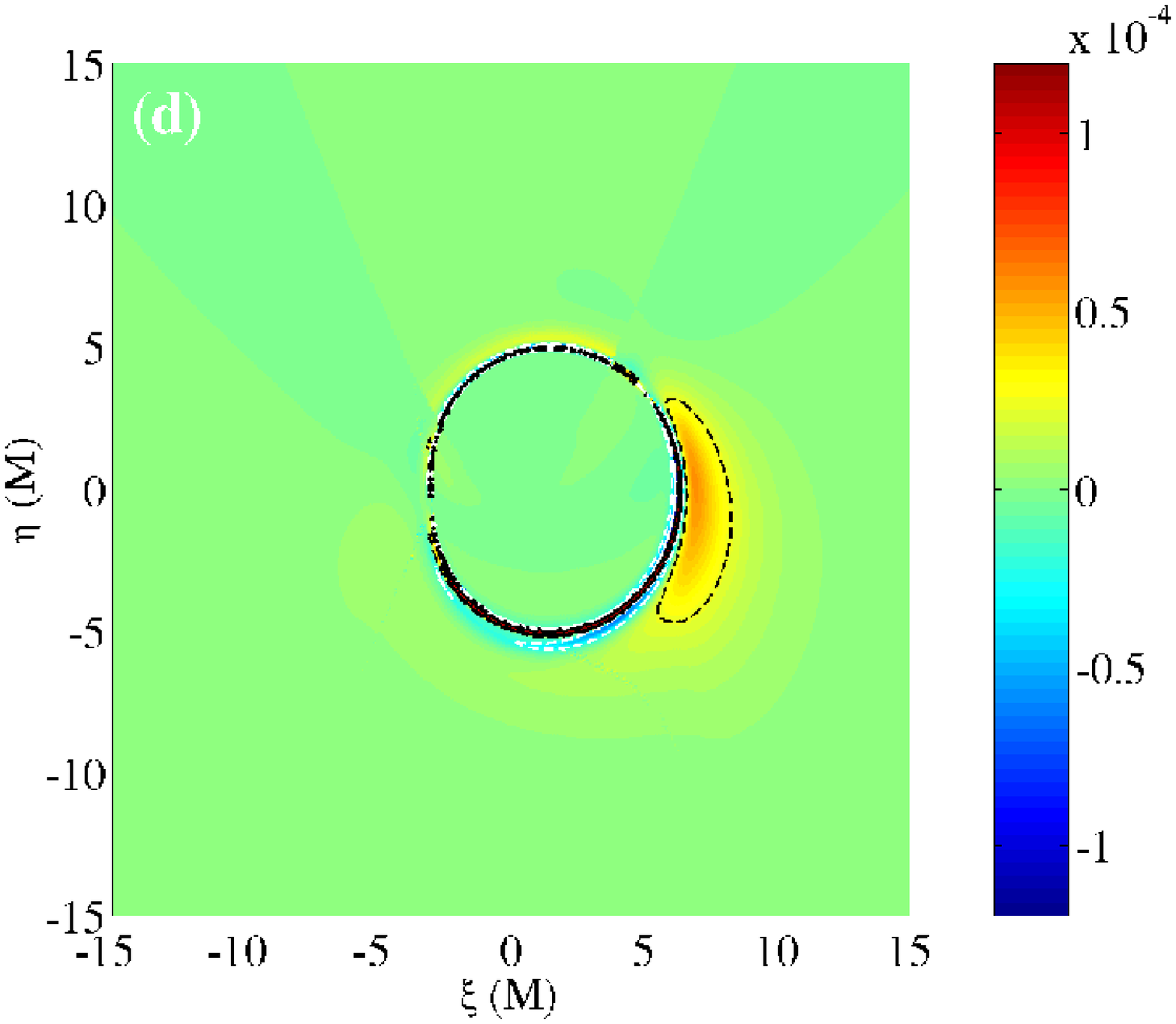}
\end{tabular}
\end{center}
\caption{
Same as Figure \ref{unpol_iquv} except with
$\omega_\infty = 3 \omega_{P\,{\rm max}}$.  The integrated fluxes over the
region shown are $I=1.0$, $Q=-4.8\times10^{-6}$, $U=2.4\times10^{-7}$,
and $V=1.2\times10^{-3}$.  All fluxes are in units of
$(M/D)^{-2} \omega_{P\,{\rm max}}^2$ as discussed above equation
(\ref{iquv_normalisation}).
}
\label{unpol_iquv_hnu}
\end{figure*}

Figure \ref{unpol_iquv} may be compared with Figure \ref{unpol_iquv_hnu}
in which Stokes $I$, $Q$, $U$, and $V$ are shown at
$\omega_\infty = 3\omega_{P\,{\rm max}}$ for the same unpolarised emission
model.  In the latter case the refractive effects are significantly
repressed.  This demonstrates the particularly limited nature of the
frequency regime in which these types effects can be expected to occur.
In this case there still does exist a net circular polarisation, now with
integrated values $I=1.0$, $Q=-4.8\times10^{-6}$,
$U=2.4\times10^{-7}$, and $V=1.2\times10^{-3}$.

\subsubsection{Polarised Emission}
\label{R:PM:PE}
\begin{figure*}
\begin{center}
\begin{tabular}{cc}
\includegraphics[width=\columnwidth]{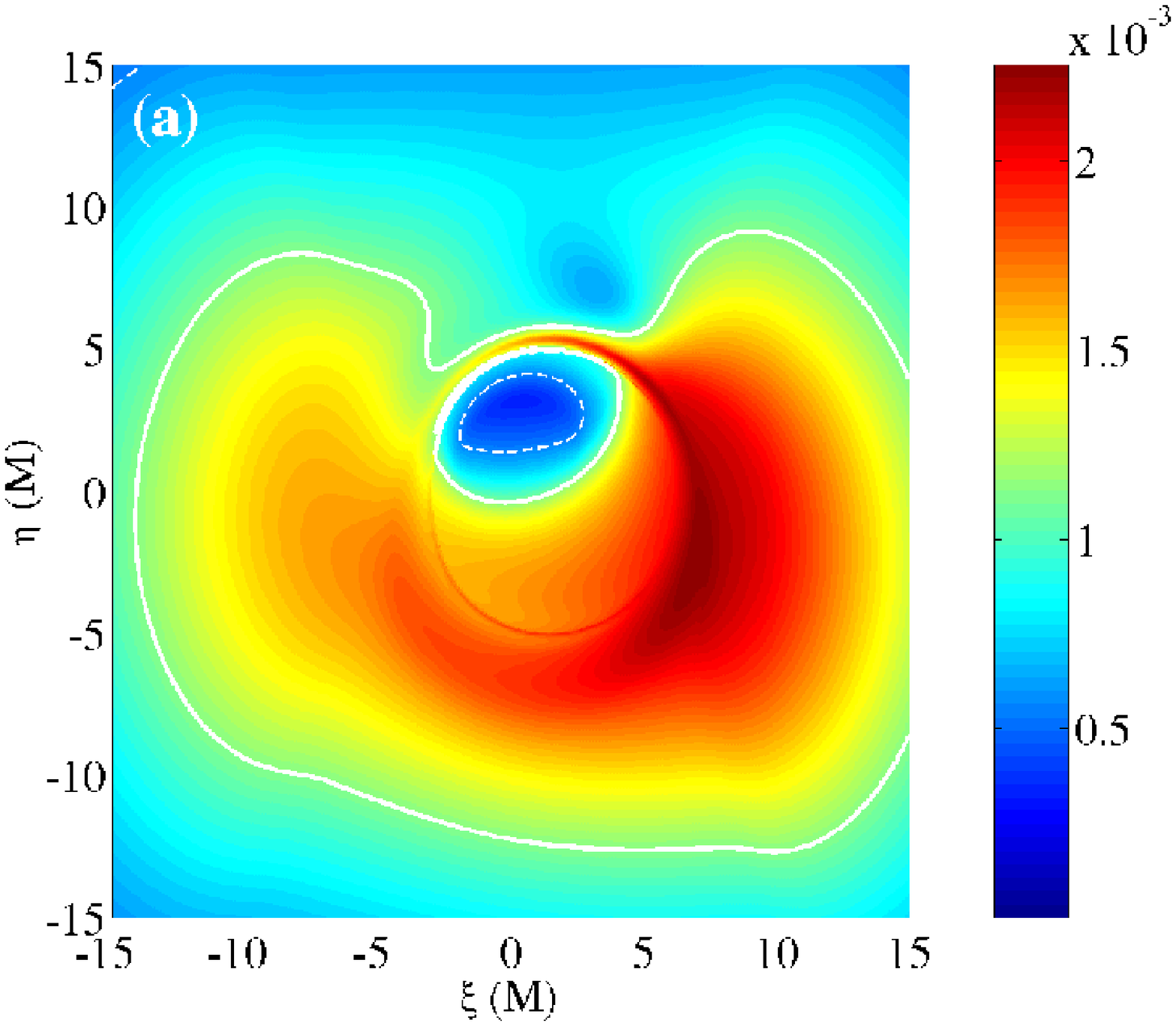}
&
\includegraphics[width=\columnwidth]{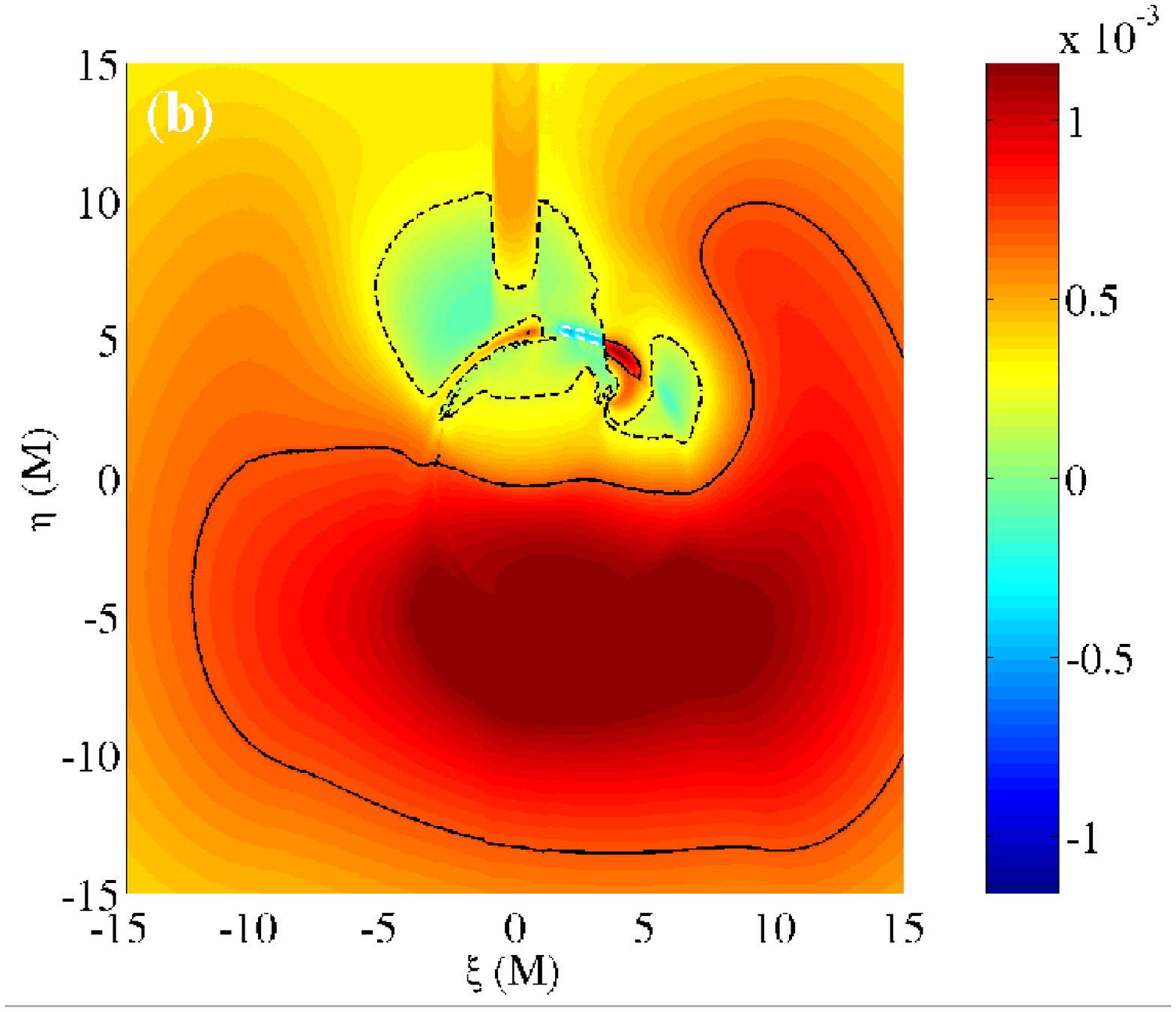} \\
\includegraphics[width=\columnwidth]{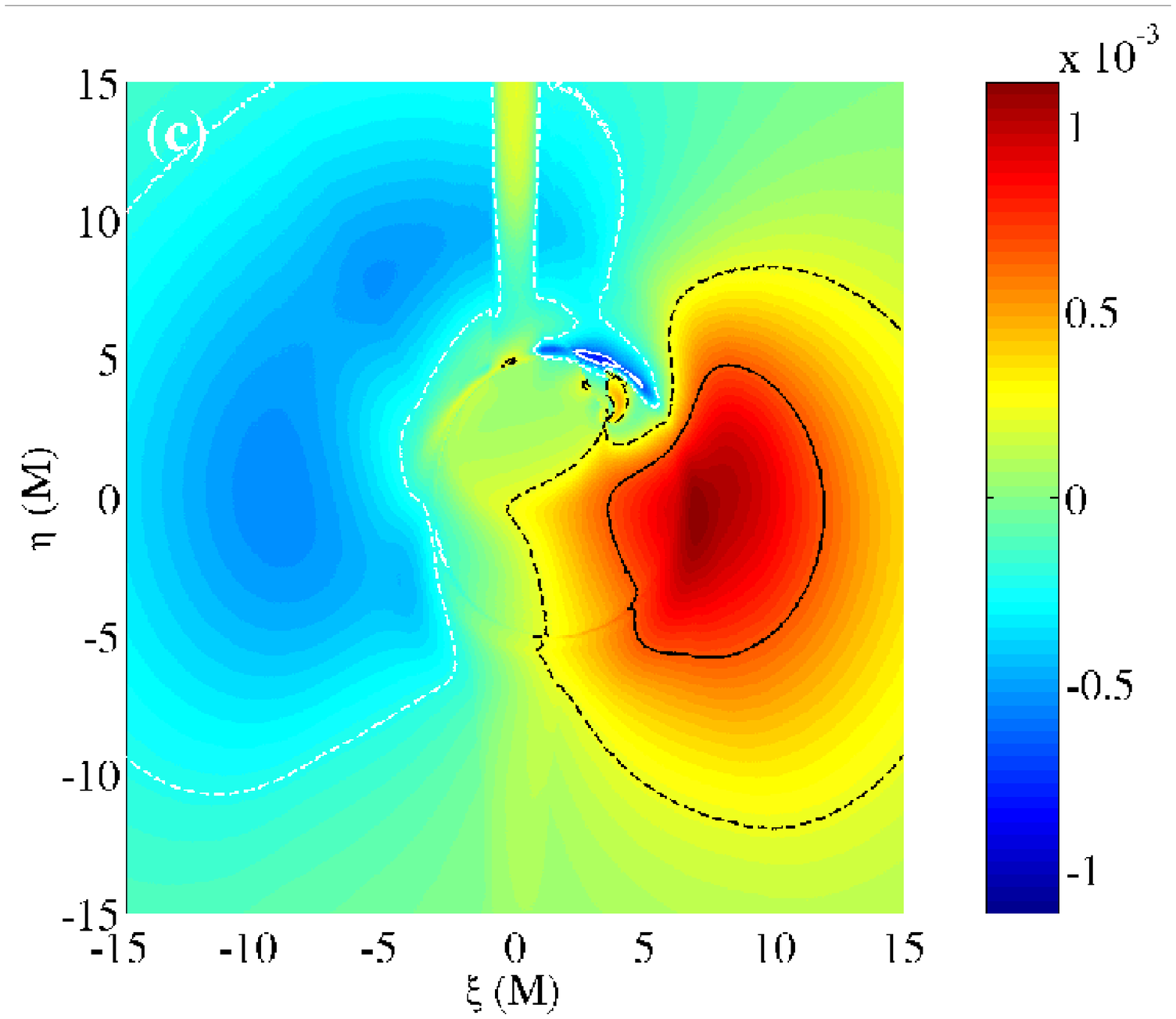}
&
\includegraphics[width=\columnwidth]{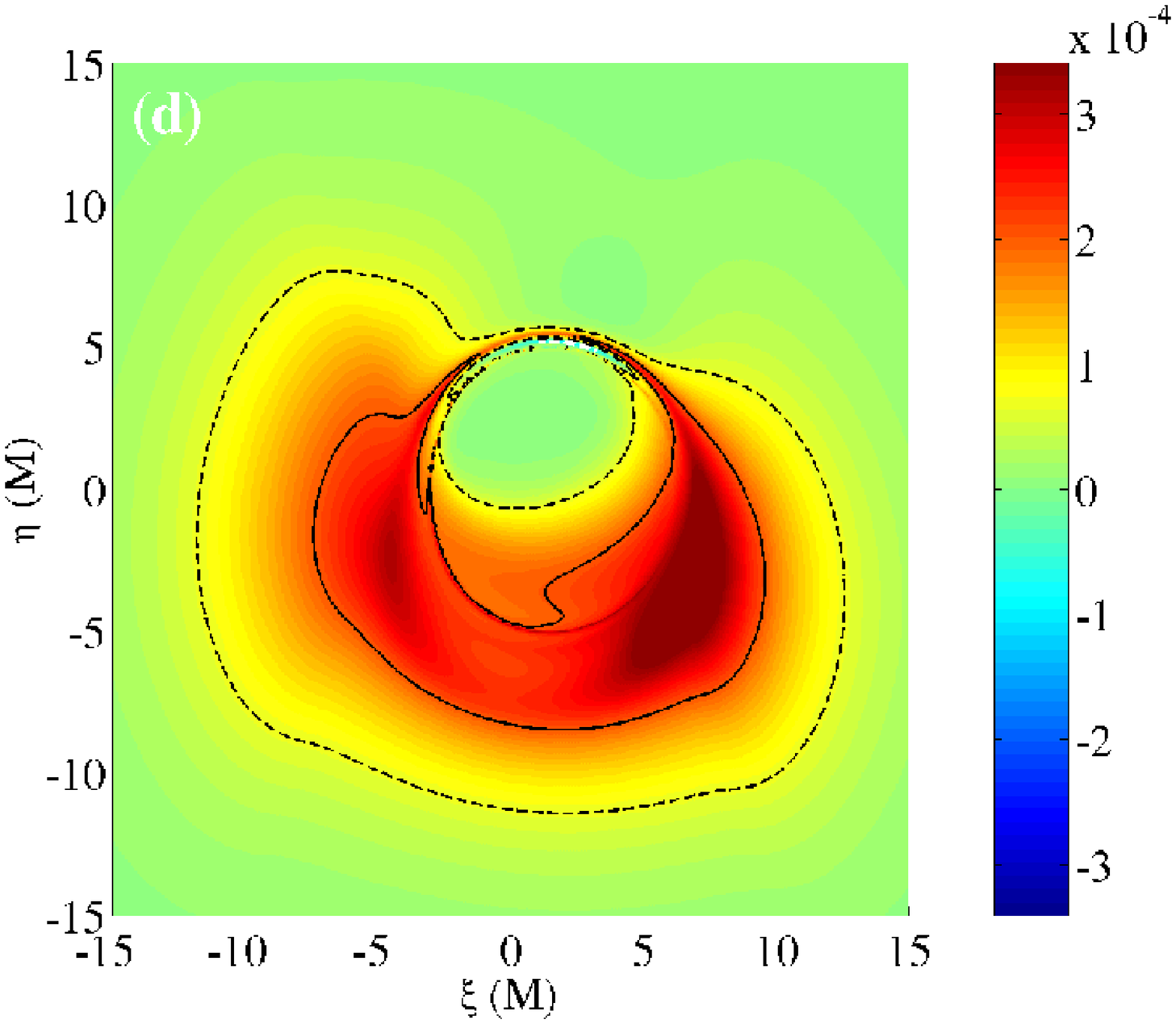}
\end{tabular}
\end{center}
\caption{
Same as Figure \ref{unpol_iquv} except using the polarised emission
mechanism (described in sections \ref{LHSRiCP:E} and \ref{LHSRiCP:AC})
and ignoring refractive plasma effects.  The integrated fluxes over the
region shown are $I=1.1$, $Q=6.0\times10^{-1}$, $U=-4.9\times10^{-3}$,
and $V=6.9\times10^{-2}$. All fluxes are in units of
$(M/D)^{-2} \omega_{P\,{\rm max}}^2$ as discussed above equation
(\ref{iquv_normalisation}).
}
\label{pol_iquv_nr}
\end{figure*}

In general, synchrotron emission will be polarised.  As a result it
is necessary to produce polarisation maps using the emission model
described in sections \ref{LHSRiCP:E} and \ref{LHSRiCP:AC}.  In this
case a net polarisation will exist even in the absence of any refraction.
In order to compare the amount of polarisation generated by refractive
effects to that created intrinsically, Figure \ref{pol_iquv_nr} shows
Stokes $I$, $Q$, $U$, and $V$ calculated using the polarised emission
model and ignoring refraction (\ie~setting the rays to be null geodesics)
for $\omega_\infty=3\omega_{P,{\rm max}}/4$.
Strictly speaking, this is a substantial
over estimate of the polarisation.  This is because, in the absence of
refraction, in principle it is necessary to include Faraday rotation
and conversion in the transfer effects considered.  As a result of the
high plasma density and magnetic field strengths, the Faraday rotation
and conversion depths for this system should be tremendous for non-refractive
rays, effectively depolarising any emission.

In comparison to Figures \ref{unpol_iquv} and \ref{unpol_iquv_hnu}, the 
general morphology of the polarisation maps are substantially different.  In
addition, the amount of linear polarisation is significantly larger, having
an integrated value of over 60\% compared to less than 0.1\% in Figure
\ref{unpol_iquv} and less than $10^{-3}$\% in Figure \ref{unpol_iquv_hnu}.
This calculation can be compared to that done by \citealp{Brom-Meli-Liu:01}.
In both it was assumed that the rays were null geodesics.  In both
Faraday rotation/conversion were neglected (in \citealp{Brom-Meli-Liu:01}
because for their disk model it was assumed to be negligible.)  However,
in \citealp{Brom-Meli-Liu:01} it was also assumed that the radiative transfer
could always be done in the adiabatic regime.  As a result, the net
polarisation was determined entirely by the emission mechanism.  However,
as discussed in section \ref{PRTiRP:LSaR} this is only possible in the
strongly coupled regime.  In this case, the dichroic terms in equation
(\ref{pol_abs}) provide the source of circular polarisation, even in the
absence of a circularly polarised emission, resulting from the different
absorption properties of the two polarisation eigenmodes.  This is
what leads to the presence of circular polarisation in Figure
\ref{pol_iquv_nr} but not in \citealp{Brom-Meli-Liu:01}.  In this case, the
integrated values of the Stokes parameters are $I=1.1$,
$Q=6.0\times10^{-1}$, $U=-4.9\times10^{-3}$, and $V=6.9\times10^{-2}$.
The vertical feature
directly above the black hole in panels (b) and (c) are associated with
the rapid decrease in the magnetic field strength in the evacuted funnel
above and below the black hole and are due to the geometric transfer
effect discussed in section \ref{PRTiRP:SCR}.

\begin{figure*}
\begin{center}
\begin{tabular}{cc}
\includegraphics[width=\columnwidth]{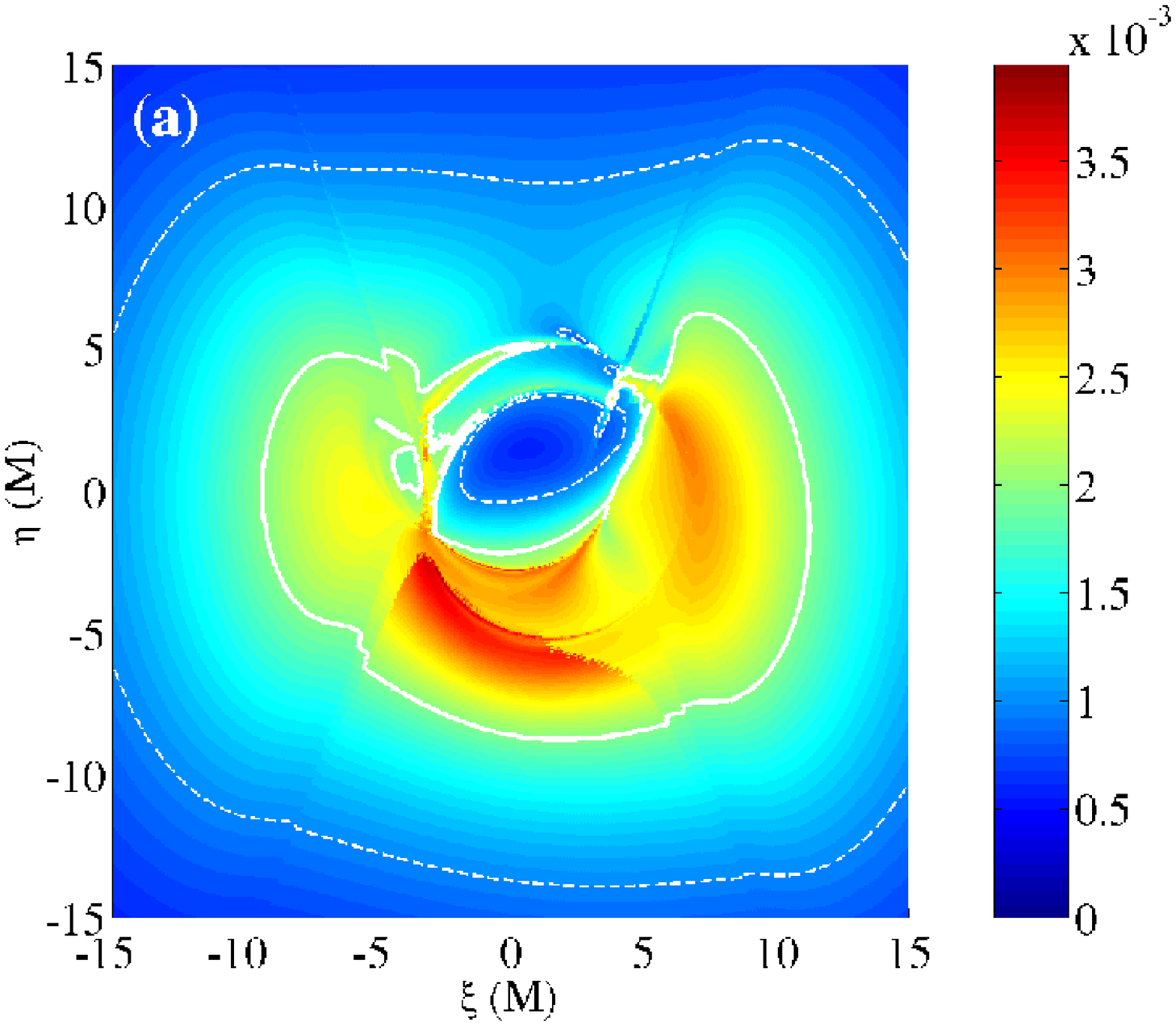}
&
\includegraphics[width=\columnwidth]{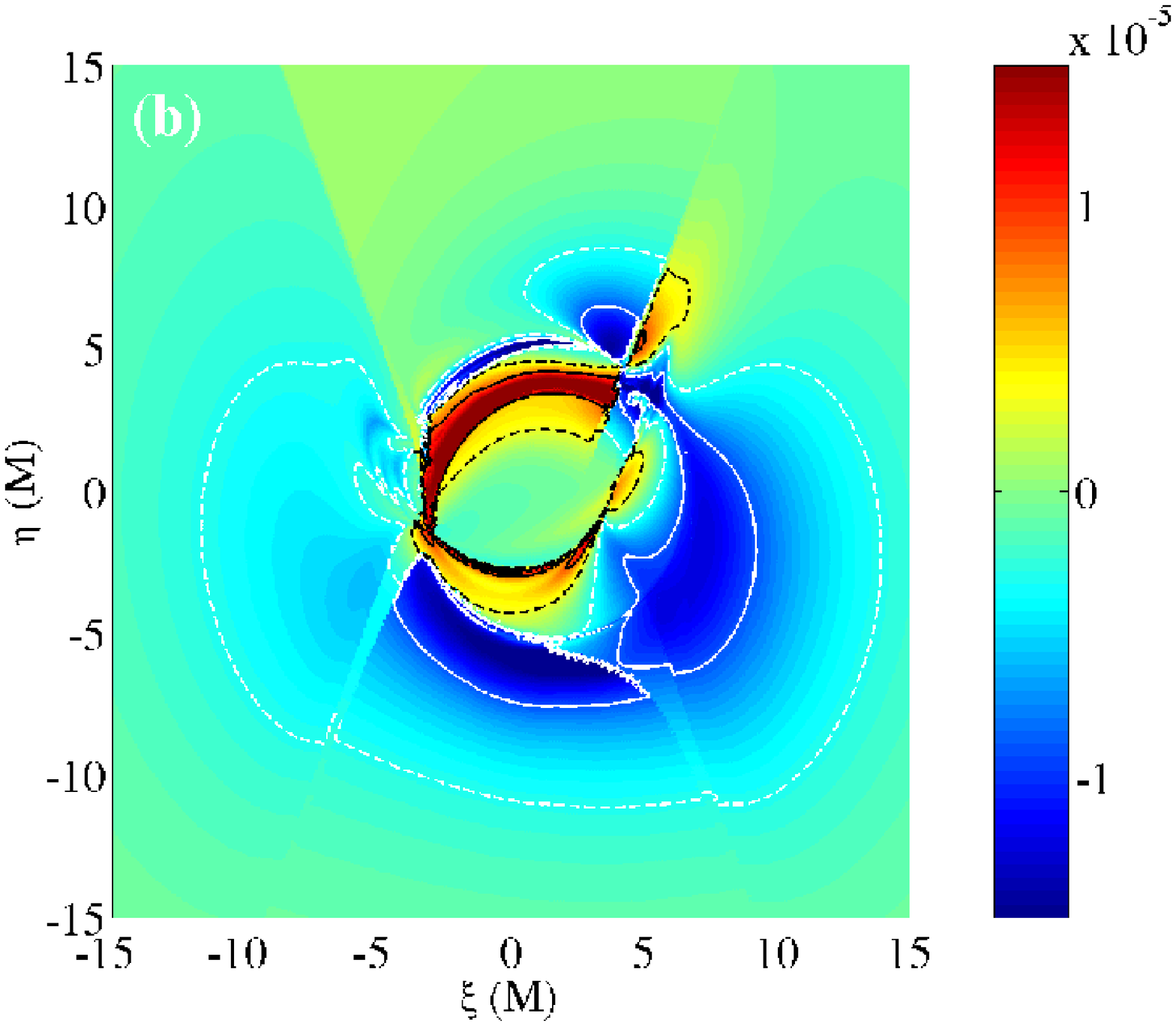} \\
\includegraphics[width=\columnwidth]{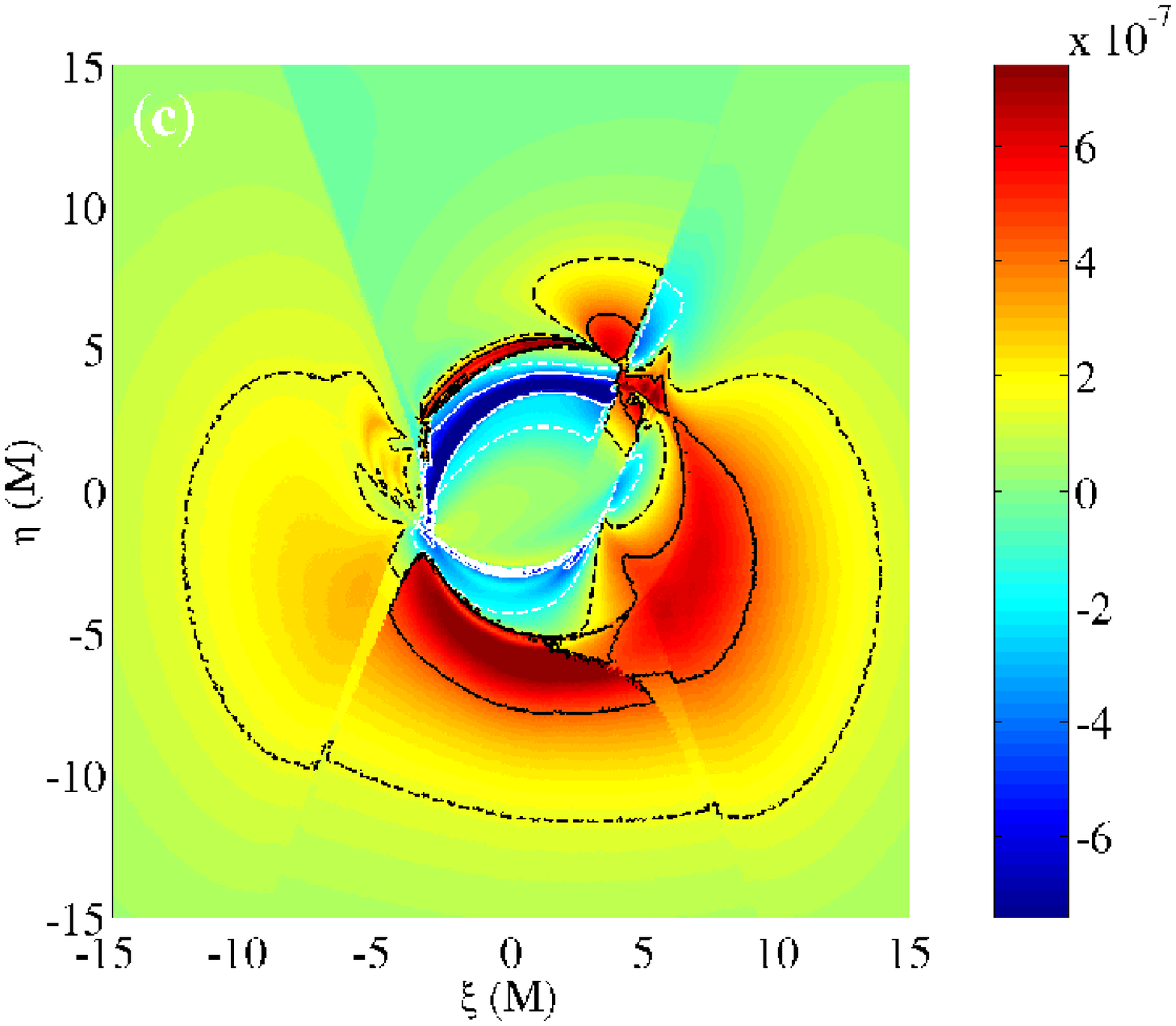}
&
\includegraphics[width=\columnwidth]{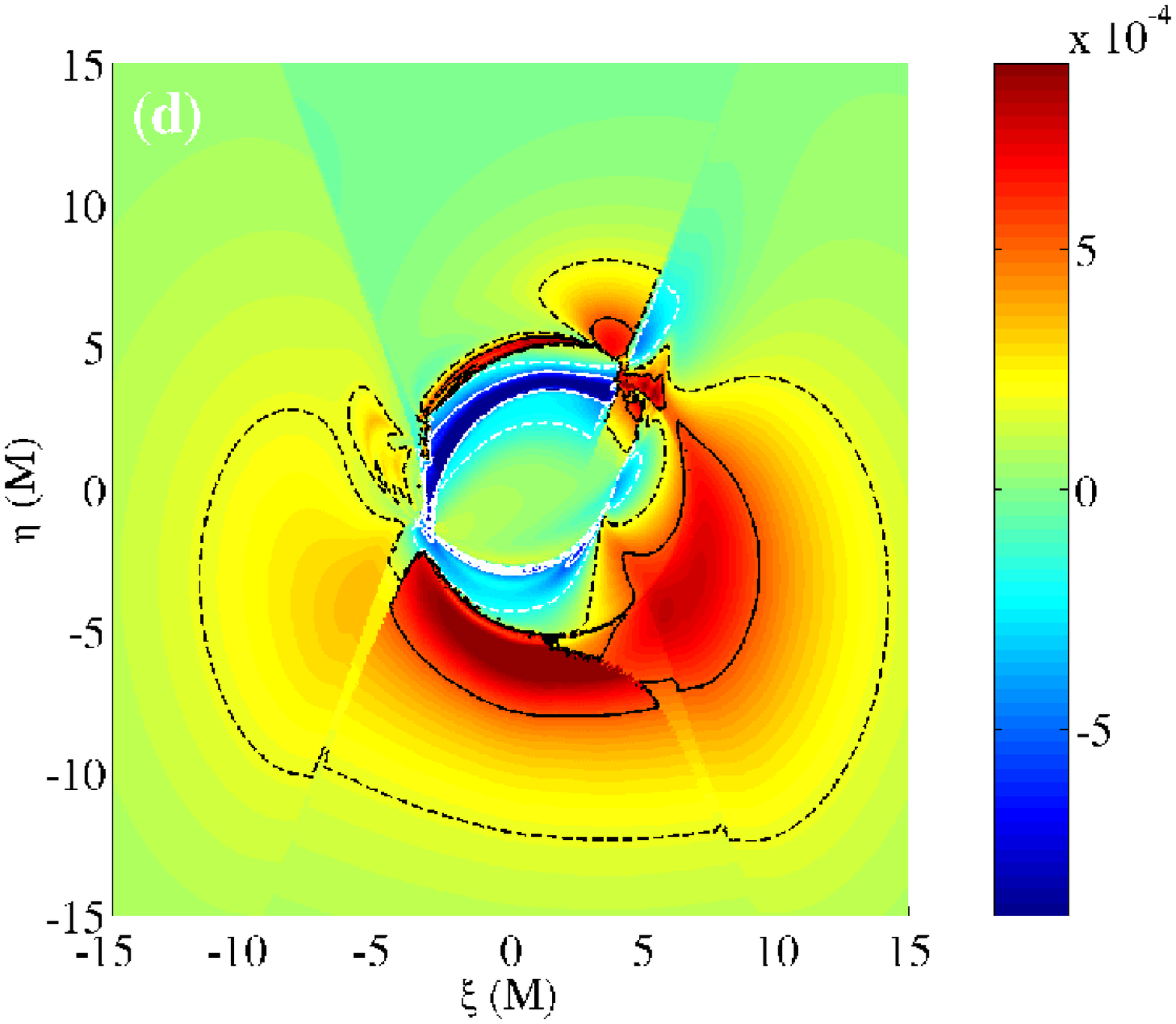}
\end{tabular}
\end{center}
\caption{
Same as Figure \ref{pol_iquv_nr} except including refractive plasma effects.
The integrated fluxes over the region shown are $I=1.3$, $Q=-2.2\times10^{-3}$,
$U=1.2\times10^{-4}$, and $V=1.4\times10^{-1}$.  All fluxes are in units of
$(M/D)^{-2} \omega_{P\,{\rm max}}^2$ as discussed above equation
(\ref{iquv_normalisation}).
}
\label{pol_iquv}
\end{figure*}

Finally, in Figure \ref{pol_iquv}, both refractive effects and the polarised
emission mechanism are included (again at
$\omega_\infty=3\omega_{P,{\rm max}}/4$).  Many of the qualitative features
of Figure \ref{unpol_iquv} still persist.  The  integrated values of the
Stokes parameters are $I=1.3$, $Q=-2.2\times10^{-3}$,
$U=1.2\times10^{-4}$, and $V=1.4\times10^{-1}$.  While the intrinsic
polarisation in the emission does make a quantitative difference, it
is clear that in this case the generic polarimetric properties are
dominated by the refractive properties.  This is most clearly
demonstrated by noting the strong supression of linear polarisation.
In Figure \ref{pol_iquv} the linear polarisation fraction is less than
0.2\% as compared with nearly 60\% in Figure \ref{pol_iquv_nr}.

\subsection{Integrated Polarisations}
\label{R:PM:IP}

\begin{figure}
\begin{center}
\includegraphics[width=\columnwidth]{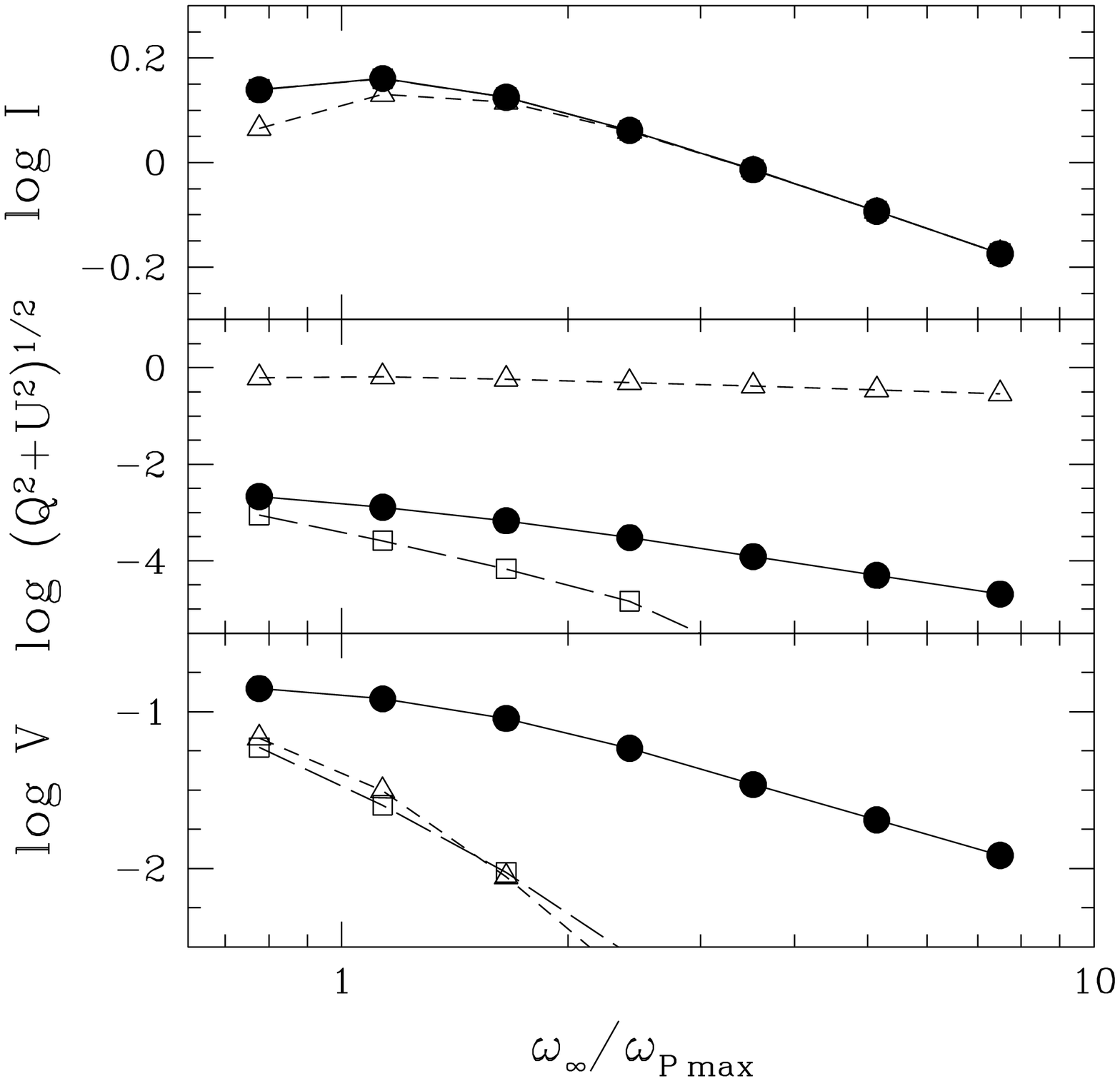}
\end{center}
\caption{The log of the integrated intensity, total linear polarisation,
and circular polarisation are shown as a function of the observation
frequency at infinity for when only polarised emission is considered
(open triangles), only refractive plasma effects are considered
(open squares), and when both are considered (filled circles).  As in
Figures 1-\ref{pol_iquv}, the disk model described in section
\ref{R:DM} and appendix \ref{aTDM} orbiting a maximally rotating black
hole is viewed from a vantage point $45^\circ$ above the equatorial plane.
  All fluxes are in units of $(M/D)^{-2} \omega_{P\,{\rm max}}^2$ as discussed
above equation (\ref{iquv_normalisation}).
}
\label{Stokes_v_nu}
\end{figure}

Figure \ref{Stokes_v_nu} shows the Stokes parameters as a function of
frequency for when
only polarised emission is considered, only refractive plasma effects are
considered, and when both are considered.  There are two notable effects
due to refraction: ({\em i}) the significant suppression of the linear
polarisation, and ({\em ii}) the large amplification of circular polarisation.
The linear polarisation is decreased by at least two orders of magnitude,
and in particular, at least two orders of magnitude less than the final
circular polarisation.  On the other hand, the circular polarisation is
more than doubled at its peak, and increases by many orders of magnitude
at higher frequencies.  Nonetheless, by
$\omega_\infty = 10 \omega_{P\,{\rm max}}$, both polarisations are
less than one tenth of their maxima.  As a result, it is clear that this
mechanism is restricted to approximately one decade in frequency, centred
about $\omega_{P\,{\rm max}}$.

\begin{figure}
\begin{center}
\includegraphics[width=\columnwidth]{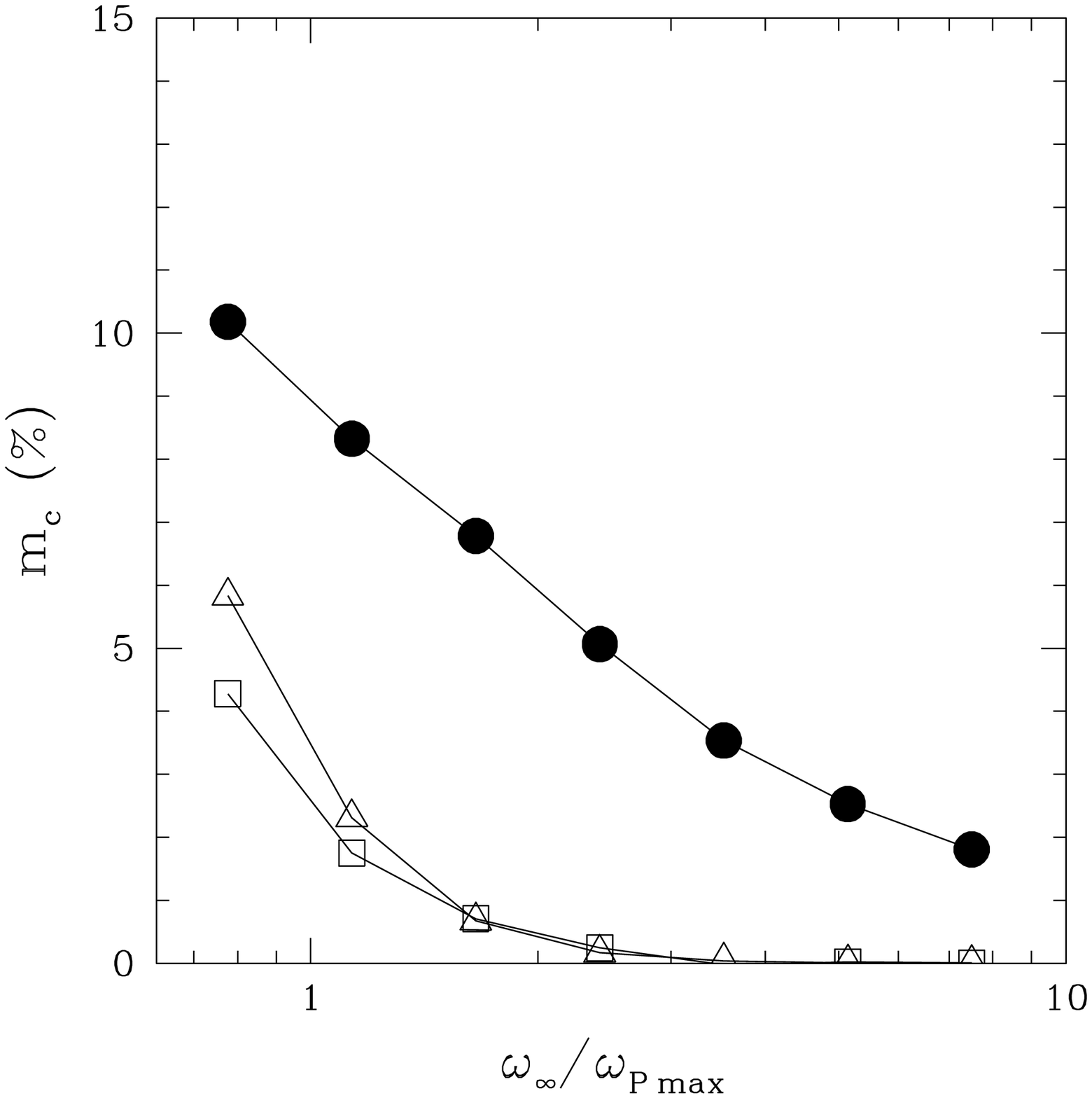}
\end{center}
\caption{Shown is the circular polarisation fraction as a function of the
observation frequency at infinity for when only polarised emission is
considered (open triangles), only refractive plasma effects are considered
(open squares), and when both are considered (filled circles).  As in
Figures 1-\ref{Stokes_v_nu}, the disk model described in section
\ref{R:DM} and appendix \ref{aTDM} orbiting a maximally rotating black
hole is viewed from a vantage point $45^\circ$ above the equatorial plane.}
\label{mc_v_nu}
\end{figure}

Figure \ref{mc_v_nu} shows the circular polarisation fraction as a function
of frequency for the same set of cases that were depicted in the previous
figure. As can be seen in Figure \ref{Stokes_v_nu}, the circular and
linear polarisation spectral index are approximately equal, and both are
softer than that of the total intensity.  The result is a decreasing
circular polarisation fraction with increasing frequency.  

\section{Conclusions}
\label{C}
We have presented refraction as a mechanism for the generation of polarisation
when $\omega_\infty \sim \omega_P,\,\omega_B$.  That this will typically result
in mostly circular polarisation is a result of the fact that the polarisation
eigenmodes are significantly elliptical only when the wave-vector and
the magnetic field are within $\omega_B/\omega$ of perpendicular, which is
usually a small number near the surface where the polarisation freezes out.
In addition to producing circular polarisation, this mechanism also
significantly suppresses linear polarisation.  Because it does require
significant refraction to take place, it is necessarily limited to
approximately a decade in frequency, making it simple to identify.

As shown in section \ref{R:PM:IP}, the resulting circular polarisation will
be softer than the intensity.  However, because of optical depth effects,
as the observation frequency increases the polarimetric properties of such
a system will be dominated be increasingly smaller areas.  As a result, the
fractional variability in such a system would be expected to increase
with frequency.  Furthermore, even though the emission may arise from
a large region, the polarimetric properties will continue to be determined
by this compact area, making it possible to have variability on time scales
short in comparison to those associated with the emission region.
In addition, variability in the circular polarisation would be expected to
be correlated with variability in the integrated intensity at frequencies
where the emission is dominated by contributions from close to the horizon
(\eg~X-rays).

Possible applications to known astrophysical sources include the
Galactic Centre (at submm wavelengths) and extinct high mass
X-ray binaries (in the infrared).  These will be disscussed in
further detail in an upcoming paper.

\section*{Acknowledgements}
We would like thank Eric Agol and Yasser Rathore for a number of
useful conversations and comments regarding this work. This research has
been supported by NASA grants 5-2837 and 5-12032.

\appendix
\section{A Thick Disk Model}
\label{aTDM}
In general, the innermost portions of the accretion flow will take the
form of a thick disk.  The equation for hydrostatic equilibrium in the
limit that $\Omega \gg v_r$ is given by
\begin{equation}
\frac{\partial_\mu P}{\rho + \frac{\Gamma}{\Gamma-1} P}
=
- \partial_\mu \ln E
+ \frac{\Omega \partial_\mu L}{1-\Omega L} \,,
\label{hydro_bal}
\end{equation}
where here $\Gamma$ is the adiabatic index, $E=-\overline{u}_t$,
$\Omega=\overline{u}^\phi/\overline{u}^t$, and
$L=-\overline{u}_\phi/\overline{u}_t$ \citep{Blan-Bege:03}.  Note
that, given the metric, any two of the quantities $E$, $\Omega$, or
$L$, may be derived from the third.  Explicitly, $\Omega$ and $L$ are
related by
\begin{equation}
\Omega = \frac{g^{\phi \phi} L + g^{t \phi}}{g^{tt} + g^{t \phi} L} \,,
\end{equation}
and the condition that
$\overline{u}^\mu \overline{u}_\mu = \overline{u}^t \overline{u}_t +
\overline{u}^\phi \overline{u}_\phi = -1$ gives $E$ in terms of $\Omega$
and $L$ to be
\begin{equation}
E = \left[ - \left(g^{tt}+g^{t\phi} L \right)
 \left(1-\Omega L\right) \right]^{-1/2} \,.
\end{equation}

In principle this should be combined
with a torque balance equation which explicitly includes the mechanism for
angular momentum transport through the disk.  However, given a relationship
between any two of the quantities $E$, $\Omega$, and $L$ specifies this
automatically.  Thus the problem can be significantly simplified if such a
relationship can be obtained, presumably from the current MHD disk
simulations.

\subsection{Barotropic Disks}
\label{BD}
For a barotropic disk the left side of equation (\ref{hydro_bal}) can be
explicitly integrated to define a function $H$:
\begin{equation}
H = \int \frac{dP}{\rho(P) + \frac{\Gamma}{\Gamma-1} P} \,,
\label{H_def}
\end{equation}
which may be explicitly integrated for gases with constant $\Gamma$ to yield
\begin{equation}
H = \ln \left( 1 + \frac{\Gamma}{\Gamma-1} \frac{P}{\rho} \right) \,.
\end{equation}
Therefore, reorganising equation (\ref{hydro_bal}) gives
\begin{equation}
\partial_\mu \left( H - \ln E \right)
=
- \frac{\Omega \partial_\mu L}{1-\Omega L} \,,
\end{equation}
which in turn implies that $\Omega$ is a function of $L$ alone.  Specifying
this function allows the definition of another function $\Xi$:
\begin{equation}
\Xi = \int \frac{\Omega(L) dL}{1 - \Omega(L) L} \,.
\label{Xi_def}
\end{equation}
Using their definitions, it is possible to solve $\Omega = \Omega(L)$
for $L(x^\mu)$ and hence $\Xi(x^\mu)$.  Then $H$ and $\Xi$ are related by,
\begin{equation}
H = H_0 + \ln E - \Xi \,,
\end{equation}
which then may be inverted to yield $\rho(H_0-\ln E + \Xi)$.  Inverting
$H$ for $\rho$ then yields $\rho(x^\mu)$.  The quantity $H_0$ sets the
density scale and may itself be set by choosing $\rho$ at some point:
\begin{equation}
H_0 = H(\rho_0) - ( \ln E - \Xi )(x_0^\mu) \,.
\end{equation}

\subsubsection{Keplerian Disk}
\label{KD}
As a simple, but artificial, example of the procedure, a Keplerian
disk is briefly considered in the limit of weak gravitating
Schwarzschild black hole (\ie~$r \gg M$).  Note that this cannot be
done in flat space because in equation (\ref{hydro_bal}) the 
gravitational terms are present in the curvature only.  For a Keplerian
flow, $\Omega = \sqrt{M/(r \sin\theta)^3} \simeq M^2 L^{-3}$.  In that case
using the definition of $\Xi$ gives
\begin{align}
\Xi &= M^2 \int \frac{dL}{L^3 - M^2 L} \nonumber \\
&= \int \frac{d \ell}{\ell^3 - \ell}
= \ln \sqrt{1 - \ell^{-2}} \,,
\end{align}
where $\ell = L/M$.  However, $\ln E$ is given by
\begin{equation}
\ln E = - \ln \sqrt{-g^{tt} ( 1-\Omega L )}
= \ln \sqrt{1- \frac{2M}{r}}- \ln \sqrt{1-\ell^{-2}} \,,
\end{equation}
and hence,
\begin{align}
H &= H_0 - \ln E + \Xi \nonumber \\
&= H_0 - \ln \sqrt{1-\frac{2M}{r}}
+ \ln \left(1-\ell^{-2}\right) \nonumber\\
&\simeq
H_0 + \frac{M}{r} - \frac{M}{r \sin\theta} \,,
\end{align}
where $\ell = \sqrt{r\sin\theta/M}$ and the weakly gravitating
condition were used.  As expected, along the equatorial plane $H$, and
therefore $\rho$, is constant.  For points outside of the equatorial plane
pressure gradients are required to maintain hydrostatic balance.

\subsubsection{Pressure Supported Disk}
\label{SKD}
Accretion disks will in general have radial as well as vertical pressure
gradients.  Inward pressure gradients can support a stable disk inbetween
the innermost stable orbit and the photon orbits, thus decreasing the radius
of the inner edge of the disk.  Around a Schwarzschild black hole this can
bring the inner edge of the disk down to $3 M$.  In a maximally rotating Kerr
spacetime this can allow the disk to extend down nearly to the horizon.

Far from the hole, accreting matter will create outward pressure gradients.
An angular momentum profile appropriate for a Kerr hole which
goes from being super to sub-Keplerian is
\begin{align}
L(r_{\rm eq}) &= \left\{
\begin{array}{l}
\left.
\left( \sqrt{ g^{t\phi~2}_{~~,r} - g^{tt}_{~~,r} g^{\phi\phi}_{~~,r} } 
- g^{t\phi}_{~~,r} \right) g^{\phi\phi~-1}_{~~,r} \right|_{r=r_{\rm eq}} \\
\quad \quad \quad \quad {\rm if~} r_{\rm eq} < r_{\rm inner} \\
\\
c_1 M^{3/2} r_{\rm eq}^{-1} + c_2 M^{1/2}
+ l_0 \sqrt{M r_{\rm eq}} \\
\quad \quad \quad \quad {\rm otherwise}
\end{array}
\right.
\nonumber \\
\Omega(r_{\rm eq}) &=
\left.
\frac{g^{\phi \phi} L + g^{t \phi}}{g^{tt} + g^{t \phi} L}
\right|_{r=r_{\rm eq}} \,,
\end{align}
where both $L$ and $\Omega$ are parametrised in terms of the
equatorial radius, $r_{\rm eq}$.  The condition that $L$ reduces to
the angular momentum profile of a Keplerian disk for radii less than
the inner radius ensures that no pathological disk structures are
created within the photon orbit.  The constants $c_1$ and $c_2$ are
defined by the requirement that at the inner edge of the disk, $r_{\rm
inner}$, and at the density maximum, $r_{\rm max}$, the angular
momentum must equal that of the Keplerian disk.  In contrast, $l_0$ is
chosen to fix the large $r$ behaviour of the disk.  The values chosen
here were $r_{\rm inner} = 1.3 M$, $r_{\rm max} = 2 M$, and $l_0 = 0.1$.
The value of $H_0$ was set so that $H(r_{\rm eq}=100 M) = 0$, thus making
the disk extend to $r_{\rm eq}=100 M$.

In addition to defining $\Omega$ and $L$ it is necessary to define
$P(\rho)$.  Because the gas in this portion of the accretion flow is
expected to be unable to efficiently cool, $\Gamma=5/3$ was chosen.
The proportionality constant in the polytropic equation of state,
$\kappa$, is set by enforcing the ideal gas law for a given temperature
($T_0$) at a given density ($\rho_0$).  Thus,
\begin{equation}
P(\rho) = \rho_0 \frac{k T_0}{m_p} \left( \frac{\rho}{\rho_0} \right)^{5/3} \,.
\end{equation}
Note that $\rho_0$ and $T_0$ provide a density and temperature scale.  A
disk solution obtained for a given $\rho_0$ and $T_0$ may be used to
generate a disk solution for a different set of scales simply by multiplying
the density everywhere by the appropriate constant factor.

\begin{figure}
\begin{center}
\includegraphics[width=\columnwidth]{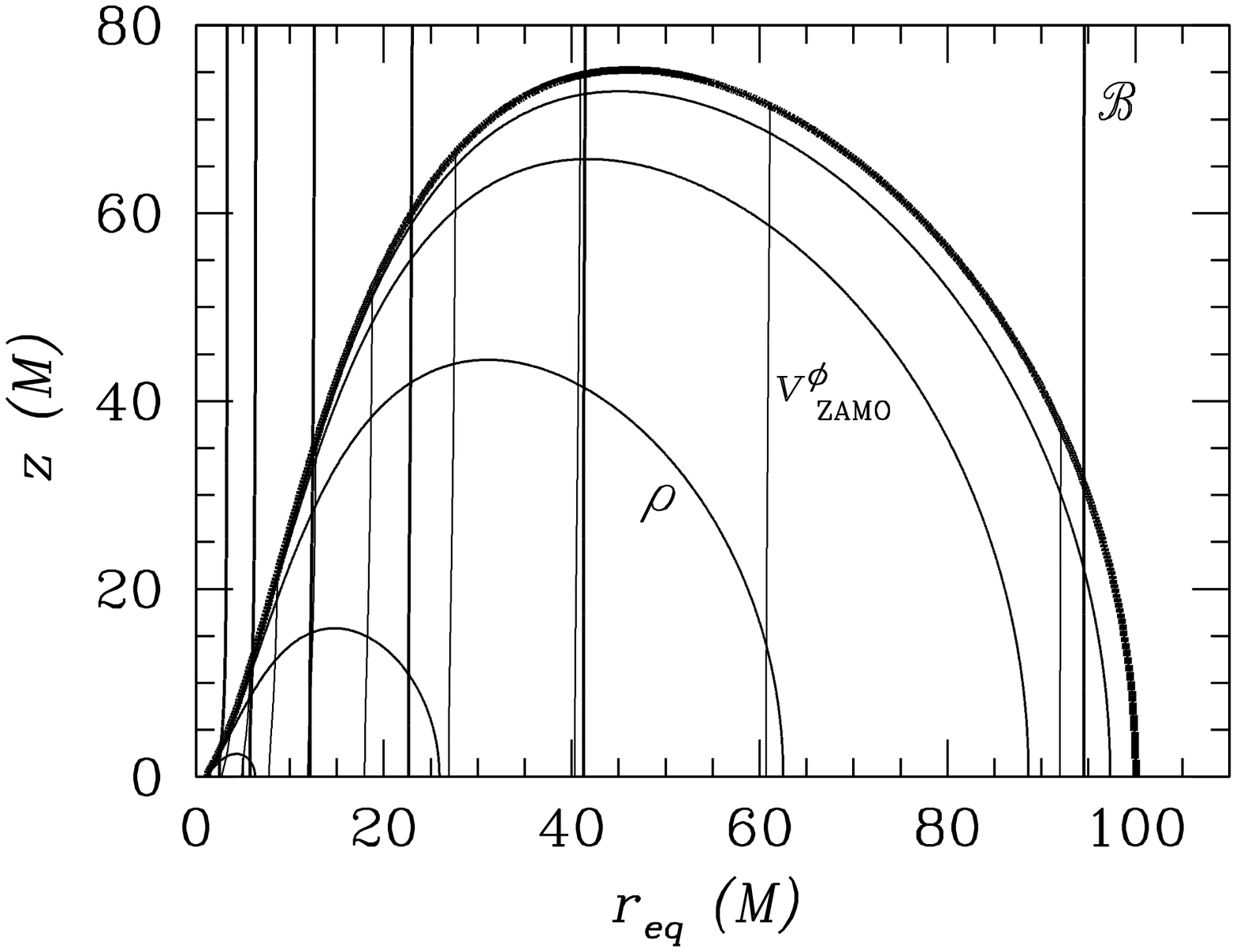}
\end{center}
\caption{Shown are the contours of the density and azimuthal velocity as
  measured by the zero angular momentum observer, and the magnetic
  field lines.
  Starting at the density maximum ($r_{\rm eq}=2M$ and $z=0$), the
  density is contoured at levels $10^{-0.5}$ to $10^{-4.5}$ times the
  maximum density in multiples of $10^{-1}$.
  From left to right, the velocity is contoured at levels $2^{-0.5} c$
  to $2^{-5} c$ in multiples of $2^{-0.5}$.  In order to provide a
  distinction between the velocity contours and the magnetic field
  lines, the velocity contours are terminated at the disks surface.
}
\label{disk}
\end{figure}

\subsection{Non-Sheared Magnetic Field Geometries}
\label{NSMFG}
The disk model discussed thus far is purely hydrodynamic.  Typically, magnetic
fields will also be present.  In general, it is necessary to perform a full
MHD calculation in order to self-consistently determine both the plasma and
magnetic field structure.  However,an approximate steady
state magnetic field can be constructed by requiring that the field lines
are not sheared.

To investigate the shearing between two nearby, space-like separated
points in the plasma, $x_1^\mu$ and $x_2^\mu$, consider the invariant
interval between them:
\begin{equation}
\Delta s^2 = \Delta x^\mu \Delta x_\mu
\quad {\rm where} \quad
\Delta x^\mu = x_2^\mu - x_1^\mu \,. 
\end{equation}
The condition that this doesn't change in the LFCR frame is equivalent to
\begin{equation}
\frac{d \Delta s^2}{d s} = 0 \,.
\end{equation}
Expanding in terms of the definition of $\Delta s$ gives,
\begin{multline}
\frac{d}{d s} g_{\mu\nu} \Delta x^\mu \Delta x^\nu
= g_{\mu\nu,\sigma} \frac{dx^\sigma}{ds} \Delta x^\mu \Delta x^\nu \\
+ 2 g_{\mu\nu} \Delta x^\mu \frac{d \Delta x^\nu}{d s} = 0\,.
\end{multline}
Note that by definition,
\begin{equation}
\frac{dx^\mu}{ds} =\overline{u}^\mu
\quad {\rm and} \quad
\frac{d \Delta x^\mu}{d s}
= \overline{u}_2^\mu - \overline{u}_2^\mu
= \overline{u}^\mu_{,\sigma} \Delta x^\sigma \,.
\end{equation}
Hence,
\begin{align}
\frac{d \Delta s^2}{d s} &= \left(
g_{\mu\nu,\sigma} \overline{u}^\sigma
+ 2 g_{\mu\sigma} \overline{u}^\sigma_{,\nu} 
\right) \Delta x^\mu \Delta x^\nu \nonumber \\
&= \left(
g_{\mu\nu,\sigma} \overline{u}^\sigma
+ 2 \overline{u}_{\mu,\nu}
- 2 g_{\mu\sigma,\nu} \overline{u}^\sigma 
\right) \Delta x^\mu \Delta x^\nu \nonumber \\
&= 2 \left(
\overline{u}_{\mu,\nu}
- \Gamma^\sigma_{\mu\nu} \overline{u}_\sigma
\right) \Delta x^\mu \Delta x^\nu \nonumber \\
&=
2 \left( \nabla_\mu \overline{u}_\nu \right) \Delta x^\mu \Delta x^\nu
 = 0 \,.
\label{no_shearing}
\end{align}
The final equality is easy to understand from a geometrical viewpoint;
for there to be no shearing, there can be no change in the direction of
$\Delta x^\mu$ of the component of the plasma four-velocity along
$\Delta x^\mu$.

That a steady state, axially symmetric magnetic field must lie upon
the non-shearing surfaces can be seen directly by considering the
covariant form of Maxwell's equations.  In particular $\nabla_\nu
^*\!F^{\mu\nu} = 0$, where $^*\!F^{\mu\nu}$ is the dual of the
electromagnetic field tensor, which in the absence of an electric
field in the frame of the plasma takes the form $^*\!F^{\mu\nu} =
{\cal B}^\mu \overline{u}^\nu - {\cal B}^\nu \overline{u}^\mu$.
Therefore,
\begin{align}
{\cal B}_\mu \nabla_\nu ^*\!F^{\mu\nu} &=
{\cal B}_\mu {\cal B}^\mu \nabla_\nu \overline{u}^\nu
+
{\cal B}_\mu \overline{u}^\nu \nabla_\nu {\cal B}^\mu \nonumber \\
& \quad
-
{\cal B}_\mu \overline{u}^\mu \nabla_\nu {\cal B}^\nu
-
{\cal B}_\mu {\cal B}^\nu \nabla_\nu \overline{u}^\mu \nonumber \\
& =
- {\cal B}^\mu {\cal B}^\nu \nabla_\nu \overline{u}_\mu = 0 \,,
\end{align}
where the first three terms vanish due to the symmetries and the
requirement that ${\cal B}^\mu \overline{u}_\mu = 0$.
This is precisely the non-shearing condition obtained in equation
(\ref{no_shearing}).

For plasma flows that are directed along the Killing vectors of the
spacetime, $\xi_i^\mu$, \ie
\begin{equation}
\overline{u}^\mu = u^t t^\mu + \sum_i u^i \xi_i^\mu \,,
\label{u_kv}
\end{equation}
where $t^\mu$ is the time-like Killing vector, it is possible to simplify
the no-shear condition considerably.
\begin{align}
&\Delta x^\mu \Delta x_\nu \nabla_\mu \overline{u}^\nu \nonumber \\
&\quad =
\Delta x^\mu \Delta x_\nu \left( u^t \nabla_\mu t^\nu
+ \sum_i u^i \nabla_\mu \xi_i^\nu \right) \nonumber \\
&\quad\quad
+ \Delta x^\mu \Delta x_\nu \left( t^\nu \partial_\mu u^t
+ \sum_i \xi_i^\nu \partial_\mu u^i \right) \nonumber \\
&\quad = 
\Delta x_t \Delta x^\mu \partial_\mu u^t
+ \sum_i \Delta x_i \Delta x^\mu \partial_\mu u^i = 0
 \,,
\label{no_shear_kv}
\end{align}
Where terms in the first parentheses vanish due to Killing's equation.  The
additional constraint that $\Delta x_\mu \overline{u}^\mu = 0$ gives
\begin{equation}
\Delta x_t = -\sum_i \Omega_i \Delta x_i \,,
\end{equation}
where $\Omega_i \equiv u^i/u^t$ is a generalisation
of the definition of $\Omega$ at the beginning of the section.  Inserting
this into equation (\ref{no_shear_kv}) and simplifying yields
\begin{equation}
\sum_i \Delta x_i \Delta x^\mu \partial_\mu \Omega_i = 0 \,,
\end{equation}
\ie~the no shear hypersurfaces are those upon which all of the $\Omega_i$ are
constant.

For the plasma flows considered in \S\ref{BD} the plasma velocity is in the
form of equation (\ref{u_kv}) where the space-like Killing vector is that
associated with the axial symmetry, $\phi^\mu$.  Thus with
$\Omega_\phi = \Omega$, the no-shear condition for this class of plasma flows
is
\begin{equation}
\Delta x^\mu \partial_\mu \Omega = 0 \,.
\end{equation}
Note that while we have been considering only axially symmetric plasma flows,
this no shear condition is more generally valid, extending to the case where
$\Omega$ is a function of $t$ and $\phi$ as well as $r$ and $\theta$.  However,
in this case it is not the perfect-MHD limit of Maxwell's equations.

For a cylindrically symmetric disk, the no-shear condition
may be used to explicitly construct the non-shearing poloidal
magnetic fields by setting
\begin{equation}
{\cal B}^r = {\cal B} \Omega_{,\theta}
\quad {\rm and} \quad
{\cal B}^{\theta} = - {\cal B} \Omega_{,r}
\,.
\label{B_def}
\end{equation}
Once the magnitude of ${\cal B}^\mu$ is determined at some point along each
non-shearing surfaces (\eg~in the equatorial plane), it may be set everywhere
by $\nabla_\mu {\cal B}^\mu -
B^\mu \overline{u}^\nu \nabla_\nu \overline{u}_\mu = 0$, which comes directly
from Maxwell's equations in covariant form and $B^\mu \overline{u}_\mu = 0$.
Inserting the form in equation (\ref{B_def}) into the first term gives
\begin{align}
\nabla_\mu {\cal B}^\mu
&=
\frac{1}{\sqrt{g}} \partial_\nu \sqrt{g} {\cal B}^\nu \nonumber \\
&=
\frac{1}{\sqrt{g}} \left( \partial_r \sqrt{g} {\cal B} \Omega_{,\theta}
- \partial_{\theta} \sqrt{g} {\cal B} \Omega_{,r} \right)
\nonumber \\
&=
\frac{1}{\sqrt{g}} \left( \Omega_{,\theta} \partial_r \sqrt{g} {\cal B}
- \Omega_{,r} \partial_{\theta} \sqrt{g} {\cal B} \right)
\nonumber \\
&=
{\cal B}^\nu \partial_\nu \ln \sqrt{g} {\cal B}
\,.
\end{align}
The second term can be simplified using equation (\ref{u_kv}),
\begin{align}
{\cal B}^\mu \overline{u}^\nu \nabla_\nu \overline{u}_\mu &= 
{\cal B}^\mu \overline{u}^\nu \nabla_\nu \left(
u^t t_\mu + u^\phi \phi_\mu \right) \nonumber \\
&=
{\cal B}^\mu \overline{u}^\nu \left(
t_\mu \partial_\nu u^t
+ \phi_\mu \partial_\nu u^\phi
- u^t \nabla_\mu t_\nu
- u^\phi \nabla_\mu \phi_\nu
\right) \nonumber \\
&=
{\cal B}_t \overline{u}^\nu \partial_\nu u^t
+ {\cal B}_\phi \overline{u}^\nu \partial_\nu u^\phi
+ {\cal B}^\mu \overline{u}^\nu \left( t_\nu \partial_\mu u^t
\phi_\nu \partial_\mu u^\phi \right) \nonumber \\
&\quad+ {\cal B}^\mu \overline{u}^\nu \nabla_\mu \overline{u}_\nu
\nonumber \\
&=
{\cal B}^\mu \left( \overline{u}_t \partial_\mu u^t
+ \overline{u}_\phi \partial_\mu u^t \Omega \right)
\nonumber \\
&=
{\cal B}^\mu \left( \overline{u}_t + \Omega \overline{u}_\phi \right)
\partial_\mu u^t + \overline{u}_\phi u^t {\cal B}^\mu \partial_\mu \Omega
\nonumber \\
&=
-{\cal B}^\mu \partial_\mu \ln u^t \,,
\end{align}
where the stationarity and axially symmetry have been used in the third step
and the no-shear condition was used in the final step.  Therefore, the
magnitude ${\cal B}$ can be determined by
\begin{align}
\nabla_\mu {\cal B}^\mu -
B^\mu \overline{u}^\nu \nabla_\nu \overline{u}_\mu
&=
{\cal B}^\mu \partial_\mu \ln \sqrt{g} {\cal B}
-{\cal B}^\mu \partial_\mu \ln u^t
\nonumber \\
&= {\cal B}^\mu \partial_\mu \ln \frac{\sqrt{g} {\cal B}}{u^t} = 0
\,,
\end{align}
and hence
\begin{equation}
\frac{\sqrt{g} {\cal B}}{u^t} = {\rm constant}
\label{B_mag}
\end{equation}
along the non-shearing surfaces.  If ${\cal B}$ is given along a curve which
passes through all of the non-shearing surfaces (\eg~in the equatorial plane),
${\cal B}^\mu$ is defined everywhere through equations (\ref{B_def}) and
(\ref{B_mag}).

\subsubsection{Non-Shearing Magnetic Fields in a Cylindrical Flow}
\label{NSMFiaKD}
An example application of this formalism is a cylindrical flow in flat space.
In this case, $\Omega$ is a function of the cylindrical radius
$\varpi \equiv r \sin \theta$.  The Keplerian disk is a specific
example with $\Omega = \varpi^{-3/2}$.  The direction of the magnetic
field is determined by,
\begin{equation}
\Omega_{,r} = \frac{d \Omega}{d \varpi} \sin \theta
\quad \quad {\rm and} \quad \quad
\Omega_{,\theta} = \frac{d \Omega}{d \varpi} r \cos \theta \,.
\end{equation}
The magnitude, ${\cal B}$ is given by
\begin{equation}
\frac{r^2 \sin \theta}{\sqrt{1-r^2\sin^2\theta \Omega^2}} {\cal B} =
f(\Omega) \,,
\end{equation}
and thus
\begin{equation}
{\cal B} = \frac{1}{r} b(\varpi) \,,
\end{equation}
where the particular form of $b(\varpi)$ depends upon the particular
form of $f(\Omega)$.  Therefore,
\begin{equation}
{\cal B}^r = b(\varpi) \cos \theta
\quad \quad {\rm and} \quad \quad
{\cal B}^\theta = -b(\varpi) \frac{1}{r} \sin\theta \,,
\end{equation}
which is precisely the form of a cylindrically symmetric vertical magnetic
field.

\subsubsection{Stability to the Magneto-Rotational Instability}
A sufficiently strong non-shearing magnetic field configuration will remain
stable to the magneto-rotational instability (MRI).  The criterion for
instability to the MRI is
\begin{equation}
\left( {\bf k} \cdot {\bf v}_{\rm A} \right)^2 < - r \frac{d\Omega^2}{d r} \,,
\end{equation}
where $\bf{k}$ is the wave vector of the unstable mode and $\bf{v}_{\rm A}$ is
the Alfv\'en velocity \citep{Hawl-Balb:95}.  For a nearly vertical magnetic
field geometry, stability will be maintained if modes with wavelength less
than twice the disk height ($h$) are not unstable.  With
\begin{equation}
v_{\rm A} = \frac{B}{\sqrt{4 \pi \rho}}
= \frac{\omega_B}{\omega_P} \sqrt{ \frac{m_e}{m_P} } \, c \,,
\end{equation}
a Keplerian disk will be stable if
\begin{equation}
\frac{4 \pi}{h} \frac{\omega_B}{\omega_P} \sqrt{ \frac{m_e}{m_P} } \, c
> \sqrt{3} \left( \frac{M}{r} \right)^{3/2} \frac{c}{M} \,.
\end{equation}
A conservative criterion may be obtained by approximating $h \simeq h_0 r$
for some constant of proportionality $h_0$, hence
\begin{equation}
\frac{\omega_B}{\omega_P} \ga 6 h_0 \sqrt{\frac{M}{r}}
\simeq 0.3 \,,
\end{equation}
for $h_0\simeq 0.1$ and $r\simeq 7$ which are typical for the disk pictured
in Figure \ref{disk}.

Comparison to equipartion fields can provide some insight into how
unrestrictive the stability criterion really is.  Given
$\beta = P_{\rm gas}/P_{\rm mag}$ and the ideal gas law it is straight
forward to show that
\begin{equation}
\frac{\omega_B}{\omega_P} = \sqrt{ \frac{2 k T}{\beta m_e c^2}}
\simeq \sqrt{3 \beta^{-1} T_{10}} \,,
\end{equation}
where $T$ is the ion temperature.  Because the ion temperature in a thick disk
will typically be on the order of or exceed $10^{12}$ K, the equipartition
$\omega_B$ ($\beta=1$) will be at least an order of magnitude larger than
$\omega_P$.  As a result the field needed to stabilise the disk against
the MRI is an order of magnitude less than equipartition strength, and hence
is not physically unreasonable.

\subsubsection{Magnetic Field Model}
Considering the restriction placed upon the magnetic field strength discussed
in the previous sections, ${\cal B}$ was set such that in the equatorial
plane
\begin{equation}
\omega_B = \omega_P + \eta \left(r+10M\right)^{-5/4} \,,
\end{equation}
where the second term provides a canonical scaling at large radii.  Here
$\eta$ was chosen to be $0.01$.

\bibliographystyle{mn2e.bst} \bibliography{cmt2.bib}

\bsp

\end{document}